\documentclass[preprint,a4paper]{elsarticle}

\makeatletter
\def\ps@pprintTitle{%
	\let\@oddhead\@empty
	\let\@evenhead\@empty
	\let\@oddfoot\@empty
	\let\@evenfoot\@oddfoot
}
\makeatother

\usepackage{amsfonts,amsmath,amsthm,amstext,amssymb}
\usepackage{graphicx,color}
\usepackage{enumerate}

\usepackage{natbib}

\usepackage{multirow} 
\usepackage{rotating}

\usepackage{marginnote}
\allowdisplaybreaks

\usepackage{lineno}

\newcommand{\bra}[1]{\left[#1\right]} 
\newcommand{\prn}[1]{\left(#1\right)} 
\newcommand{\cur}[1]{\left\{#1\right\}} 
\newcommand{\abs}[1]{\left|#1\right|} 
\newcommand{\mb}[1]{\mathbf{#1}} 

\newcommand{\field}[1]{\mathbb{#1}} 
\newcommand{\N}{\field{N}} 
\newcommand{\R}{\field{R}} 
\newcommand{\B}{\mathfrak{B}} 

\newtheorem{assumption}{Assumption}
\newtheorem{corollary}{Corollary}
\newtheorem{definition}{Definition}
\newtheorem{lemma}{Lemma}
\newtheorem{proposition}{Proposition}
\newtheorem{theorem}{Theorem}
\renewenvironment{proof}{\textbf{Proof} \nopagebreak }{\hspace{\stretch{1}}\textbf{q.e.d.} \nolinebreak \rule{0.5em}{0.5em}}

\newcommand{\st}{\rule{0pt}{2.5ex}}

\newcommand{\cartes}[3][m]{\resizebox{!}{0.4cm}{$\times$}_{#1=1}^{#3}#2_{#1}} 

\newcommand{\cm}[1]{\textcolor{black} { #1}}

\newcommand{\dl}[1]{\textcolor{black}{#1}} 

\begin{document}
\begin{frontmatter}

\underline{\textbf{Preprint version submitted to Elsevier \hfill February 26, 2020}}\\

\begin{center}
	{\textbf{\Large Multi-machine preventive maintenance scheduling with imperfect interventions: A restless bandit approach}}
\end{center}	
\begin{center}
	Diego Ruiz-Hern\'andez, Jes\'us Mar\'ia Pinar-P\'erez, and David Delgado-G\'omez
\end{center}
\vspace{5mm}
\begin{center}
	{\large Published in \textbf{Computers and Operations Research} (ELSEVIER), July 2020.\\}
\end{center}
\vspace{5mm}
\textbf{Cite as:} Ruiz-Hern\'andez, D., Pinar-P\'erez, J. M., and Delgado-G\'omez, D. (2020). Multi-machine preventive maintenance scheduling with imperfect interventions: A restless bandit approach. Computers and Operations Research, 119, 104927.\\
\begin{center}
	\textbf{DOI:} https://doi.org/10.1016/j.cor.2020.104927
\end{center}
\vspace{5mm}
\begin{flushleft}
	Article available under the terms of the \textbf{CC-BY-NC-ND} licence
\end{flushleft}

\vfill
\textbf{Preprint version submitted to Elsevier \hfill February 26, 2020}

\title{Multi-machine \cm{preventive} maintenance scheduling with imperfect interventions: a restless bandit approach} 


\author[DiRu]{Diego Ruiz-Hern\'andez\corref{Diego}}
\author[JePi]{Jes\'us M. Pinar-P\'erez}
\author[DaDe]{David Delgado-G\'omez }

\cortext[Diego]{Corresponding author.  \textit{Email address}: d.ruiz-hernandez@sheffield.ac.uk}
\address[DiRu]{Sheffield University Management School,   Conduit Road, S10 1FL, Sheffield, UK}
\address[JePi]{University College for Financial Studies (CUNEF), Leonardo Prieto Castro 2, 28040, Madrid, Spain}
\address[DaDe]{Universidad Carlos III de Madrid, Av. de la Universidad 30, 28911,  Madrid, Spain}

\begin{abstract}
In this paper we address the problem of allocating the efforts of a collection of repairmen to a number of deteriorating machines in order to reduce operation costs and to mitigate the cost (and likelihood) of unexpected failures. Notwithstanding these preventive maintenance interventions are aimed at returning the machine to a so-called \textit{as-good-as-new} state, unforeseeable factors may imply that maintenance interventions are not perfect and the machine is only returned to an earlier (uncertain) state of wear. The problem is modelled as a restless bandit problem and an index policy for the sequential allocation of maintenance tasks is proposed. A series of numerical experiments shows the strong performance of the proposed policy. Moreover, the methodology is of interest in the general context of dynamic resource allocation and restless bandit problems, as well as being useful in the particular imperfect maintenance model described.
\end{abstract}


\end{frontmatter}

\section{Introduction}

Maintenance interventions in deteriorating equipment are aimed at guaranteeing its availability, reliability and safe performance whilst maximising output and minimising waste. Such interventions may be \cm{preventive}, aimed at guaranteeing an optimal performance of the equipment and preventing eventual failures and breakdowns by means of minor actions as cleaning surfaces, lubricating joints, sharpening blades, replacing and refilling fluids, removing waste, among others \cm{\cite{Percy2008}}; or corrective, aimed at repairing minor or major faults in the system. Regarding the planning horizon, whereas corrective maintenance is reactive, preventive interventions are scheduled using either time-based or condition-based schemes. Time-based interventions are usually scheduled according to an age-based regime or following a pre-determined calendar; while condition based interventions are conducted in irregular intervals based the state of wear of the equipment, which is typically determined by means of inspections or a -more or less sophisticated- monitoring mechanism. 

\cm{Typically the aim of a given maintenance intervention is returning the equipment to an \textit{as-good-as-new} state. However, it is easy to find scenarios where the quality of a maintenance intervention may be affected by uncontrolled factors associated, for example, with the skills and qualification of the technicians, the quality or suitability of the tools and components used, or the existence of hidden defects that are not typically undercover during a routine check \cite{NakYas1987,PhaWan1996}. This implies that a maintenance intervention may actually take the machine to an earlier (undetermined) state of wear, improving its current condition but without returning the machine to its pristine state. For a more in depth discussion of general machine maintenance issues see, for example, \cite{KobMur2008,MobHigWik2008,BenDufRao2009}.}

\cm{When the target of a preventive maintenance strategy is a single machine, maintenance strategies are designed either to determine appropriate intervention times, or to identify the subsystems or components that should be intervened, or both. The objective is preventing failures and guaranteeing a high level of performance at a minimum cost. However, when a system consists of several machines and only a limited number of repairmen is available, the objective of a maintenance scheduling regime is to determine which machines to intervene at each decision epoch. }

\vspace{6pt}

\cm{As it is argued in the literature review, the problem of scheduling fortuitously imperfect maintenance interventions in multiple deteriorating equipment has not yet fully addressed in academic literature. This is especially relevant when it has been highlighted that realistic and valid maintenance scheduling methods must incorporate random features of the maintenance tasks \cite{PhaWan1996}. In this article, we intend to fill this void by proposing an index heuristic based on a restless bandit formulation of the problem. Rather than a long-term scheduling plan, we propose an on-line heuristic system that schedules interventions based on information available on the equipment's condition at each decision time.  }

\cm{Notwithstanding the research is motivated by the machine maintenance problem, it is important mentioning that the proposed methodology can be seen as a generic intervention scheduling model, that can be used for systems of independent bi-directionally-evolving elements when the result of the interventions is uncertain.}

\section{\cm{Literature Review}}

\cm{There is a vast amount of literature addressing the problem of scheduling preventive maintenance interventions in one single machine. In most cases it is assumed that maintenance actions are perfect and, upon intervention, the equipment or machine is returned to an \textit{as-good-as-new} state. Some examples of work in this area, which are somehow related to our approach, are: Grall et al.  \cite{GraBerDie2002} study the problem of designing condition-based maintenance policies in one-machine systems. The machine degradation is modelled as a Markovian process. Borrero and Akhavan-Tabatabaei \cite{BorAkh2013} analyse an inventory/maintenance problem for single machine systems using a Markov Decision Process approach. Inventory holding and unfulfilled demand costs are the main variables in the decision process. Zhu et al.  \cite{ZhuPenVan2015} propose a condition-based maintenance policy for a continuous degrading multi-component system in an infinite horizon.  They develop an analytical solution suitable for large-scale problems. Poppe et al. \cite{PopBouLam2018} address a condition-based maintenance problem in a one-machine multi-component system. They propose a double-threshold based hybrid policy for scheduling preventive and corrective maintenance interventions based on the machine degradation level.}

\cm{In some other cases, a maintenance intervention is planned to be intentionally partial (i.e. conducting minor repairs or replacements instead of a thorough revision of the equipment), aimed at attaining minor improvements in the machine's performance instead of taking it to an \textit{as-good-as-new} state. This problem is typically referred to as imperfect maintenance or minimal maintenance. In all these cases, the state of the machine after intervention is assumed to be known or determined.  Kurt and Kharoufeh \cite{KurKha2010} propose threshold policies for the maintenance/replacement problem in a single equipment while degradation is driven by a discrete Markovian process. Maintenance decisions are taken following periodical inspections of the equipment. The interventions are assumed imperfect, affecting either the virtual-age or the hazard-rate function of the system. Assuming non-decreasing operating costs, the authors propose an optimal threshold-type policy based on the system deterioration state. Moghaddam and Usher \cite{MogUsh2011} use dynamic programming techniques for solving the problem of scheduling both preventive maintenance and replacement tasks in deteriorating multi-component equipment. The effectiveness of the intervention is given by a known improvement factor.  Huynh et al. \cite{HuyCasBar2012} analyse minimal repair maintenance strategies in single-unit systems subject to degradation and external shocks. Their aim is to study different approaches for conducting minimal repair (either time-based or degradation-based) as a reaction to the degrading effect of traumatic shocks; they compare the overall cost savings of these policies with the costs of a pure age-based replacement policy. Do Van et al \cite{DoVoiLev2015} compare the costs and efficiency of two alternative families of policies: perfect and imperfect maintenance. They propose a maintenance policy aimed at optimally selecting the type of maintenance action to be taken at each inspection time. The time between interventions is determined by a remaining-useful-life inspection policy. Gilardoni et al. \cite{GilTolFre2016} analyse the partial maintenance problem of deteriorating equipment and propose static and dynamic intervention policies based on information about the failure history of the equipment.  Lee and Cha \cite{LeeCha2016} address a one-machine periodic preventive maintenance problem with minimal repair. Minimal interventions are conducted periodically for improving the machine's reliability performance. Failures between maintenance interventions, which are modelled by means of a generalised non-homogeneous Poisson process, are assumed to be immediately repaired. Other relevant references are \cite{LiaPanXi2010,LiuHua2010b, NguDijFou2017}.}

\cm{The problem of single-machine maintenance interventions with uncertain outcome, to which hereby we refer to as imperfect maintenance, has been studied by a small number of authors. Wang and Sheu \cite{WanShe2003} address a multi-machine production/maintenance problem subject to inspection errors.  Liao et al. \cite{LiaElsCha2006} analyse a condition-based maintenance model for continuously deteriorating systems. It is assumed that after an intervention the equipment is returned to a randomly determined state with residual damage. The authors propose a so-called condition-based availability limit policy, aiming at maximising the availability level of the system. Meier et al. \cite{MeiRibSou2009} discuss the maintenance model of a deteriorating system where the efficiency of the interventions is modelled as a normally distributed random variable. Different maintenance strategies are evaluated and their long-term costs computed.  Do Van et al.\cite{VanBer2012} propose a condition-based maintenance policy for deteriorating equipment. The authors assume that maintenance interventions can be either deterministically imperfect (the system is intentionally taken to a pre-determined deterioration level); or randomly imperfect (the system is fortuitously  returned to a certain deterioration level). In both cases, the final state of the machine is better than the initial one. The system determines the optimal intervention time and the type of action to be taken in order to minimise a measure of maintenance cost. Mor and Mosheiov \cite{MorMos2015} address the problem of scheudling maintenance interventions in single-machine system where delays in the intervention increase the completion time of the task. Two deterioration regimes, time and state dependent, are discussed. They demonstrate that their formulation can be reduced to a linear assignment problem. Khatab and Aghezzaf \cite{KhaAgh2016} address the problem of selective stochastic maintenance scheduling in a multi-component system, where interventions can be conducted during specific equipment's idle times. Different actions ranking from minimal repair to replacement are available at each intervention epoch. The quality of a maintenance action is assumed to be stochastic (with known probability distribution), depending on the skills of the repair team. Their objective is determining the optimal subset of tasks (cost minimising) to be conducted during the idle times in order to guarantee optimal system's performance in subsequent stages. Liu et al.  \cite{liuWuXie2017} propose a maintenance policy for deteriorating equipment with age- and state-dependent operating costs. The authors develop a replacement model that is further extended to a repair/replacement model. Partial maintenance interventions are aimed at returning the system to certain operating level, but the outcome can be controlled or not.  They establish the optimality of a monotone control-limit policy. }

\cm{Regarding literature addressing the problem of scheduling maintenance interventions to more than one machine, we notice that most of these works assume that the outcome or quality of the maintenance interventions is perfect or, at least, deterministic. Among them, Kenn\'e et al. \cite{KenBouGha2003} provide an algorithm for solving an optimal control problem consisting of finding optimal -cost minimising- repair rates for a collection of identical independent machines subject to breakdowns. Gharbi and Kenn\'e \cite{GhaKen2005} apply control theory and simulation techniques to find production and preventive maintenance rates for a multiple-machine manufacturing system in order to minimise total inventory, repair and maintenance cost. Wang and Pham \cite{WanPha2006b} discuss models that, in general, assume equipment dependency and interaction and focuses in group and opportunistic policies. Ruiz, Garc\'ia-D\'iaz, and Maroto \cite{RuiGarMar2007} propose a formulation that simulataneously schedules a series of jobs together a preventive maintenance intervention. The aim is minimising the makespan of the complete sequence.  When addressing the multi-machine maintenance problem, P\'erez-Canto \cite{Canto2008} assumes that the number of machines is limited and smaller than the number of technicians, and the schedule tableau is designed beforehand.  Oyarbide-Zubillaga et al. \cite{OyaGotSan2008}  address the problem of determining the optimal frequency of preventive maintenance interventions for multiple-machine systems. They use discrete event simulation and evolutionary algorithms for solving instances with different costs and profit criteria. Mosheiov and Sarig \cite{MosSar2009} propose a heuristic for the problem of scheduling maintenance activities in independently deteriorating machines. Wang and Wei \cite{WanWei2011} address the problem of scheduling maintenance activities in a collection of parallel deteriorating machines in order to either minimise differences in job completion times, or total differences in waiting times. Hsu et al. \cite{HsuJiGuo2013} address the problem of scheduling one single maintenance intervention on a collection of parallel aging machines in order to minimise total completion time and total machine load. Moghaddam \cite{Mog2013} proposes a multi-objective non-linear mixed integer optimization model to identify Pareto-optimal preventive maintenance and replacement schedules for a multiple-machine manufacturing system. More recently, Tao et al. \cite{TaoXiaXi2014} deploy a theory of constraint-based models in order to find a savings maximising regime for scheduling opportunistic preventive maintenance interventions.  Irawan et al. \cite{IraOueJon2017} deal with the problem of scheduling maintenance when the recommended intervention period is given. The problem is set within the context of a wind farm. Also within the wind energy context, Froger et al. \cite{FroGenMen2017} propose a branch-and-check heuristic for finding efficient solutions to a multi-equipment maintenance scheduling problem. Seif and Yu \cite{SeiYu2018} address a general operation and maintenance planning problem for multi-machine systems. The problem is formulated as a MIP problem and a solution method is provided.  Liu et al. \cite{liuChenJiang2018}  propose a selective maintenance policy, choosing the sequence of actions to be taken in an integrated system (with dependencies). Detti et al. \cite{DetNicPac2019} address the problem of secheduling a collection of jobs and a manitenance activity in a single-machine system. They propose a mixed-integer programming formulation of the problem. Good reviews of contributions to single and multi-machine maintenance management and optimisation can be found in \cite{PhaWan1996,Wan2002,AlaXia2017,CaoJiaHu2018}. }

\cm{Among the techniques proposed for the scheduling of preventive maintenance interventions in independently deteriorating equipment stands-out, due to its conceptual simplicity and outstanding performance, the family of so-called index policies. Such policies root from the classical result of Gittins \cite{Gittins} and Whitte \cite{Whit1} for the multi-armed bandit problem, and its extension to the restless bandit problem by Whittle \cite{Whit3} among many others (a thorough account of the evolution of this field can be found in Glazebrook et al. \cite{GlaHoKi2014}). For example, Whittle \cite{WhitOC}, Glazebrook \cite{GlaMitAns}, and Abad et al. \cite{AbIy2016} propose index policies for solving particular maintenance problems; likewise, Glazebrook et al. \cite{GlaRuKi} establish general indexability for a family of problems where maintenance interventions have to be scheduled in order to mitigate increasing operation (and breakdown) costs due to machine deterioration. However, to our knowledge, this methodology has not yet been applied to a framework where the outcome of a maintenance intervention is uncertain.}

\dl{Among the multiple -exact and heuristic- techniques that have been proposed for the scheduling of preventative maintenance interventions in large systems of independently deteriorating equipment stands-out, mostly because of its conceptual simplicity but also because of its outstanding performance, the family of so-called index policies. Such policies root from the classical result of Gittins \cite{Gittins,GitJon,GitGlaWe2011} and its further refinements by Whittle \cite{Whit1,Whit2,Whit3} and many others (for a thorough account of the evolution of this area please refer to the work by Glazebrook et. al. \cite{GlaHoKi2014})}.

\dl{Roughly speaking, an index heuristic consists of calibrating an state-de\-pen\-dent index for each machine/state combination in the system, and allocating -at each decision epoch- the effort to the subset of machines with largest indices associated to their current states. Gittins \cite{Gittins} (and later Pandelis and Teneketzis \cite{PanTen}) proved that, in a setting where all the elements in the system that are not been intervened remain in their current state, the index policy which allocates the effort to the machine or machines with larger index is, indeed, optimal. Whittle \cite{Whit3} extended these results to the case where those elements that are not receiving effort are allowed to evolve over time. Notwithstanding this class of problems has been shown to be intractable \cite{PapaTsi}, a large number of empirical studies have witnessed its outstanding performance in different areas of application \cite{AbIy2016,KG02,GlaMitAns}. Moreover, Weber and Weiss \cite{WebWei} established a form of asymptotic optimality for Whittle's index heuristic and Glazebrook et. al. \cite{KG01} developed certain bounds on its suboptimality.}

\vspace{6pt}

\cm{In this article we address the problem of scheduling (fortuitously imperfect) preventive maintenance interventions to a collection of independent deteriorating machines. We assume that the number of repairmen (or teams) is smaller than the number of machines and, therefore, a sound effort allocation policy has to be developed in order to minimise operation/intervention costs and the risks of failures due to machine wear. Unfortunately, the size of real life instances prevents the direct aplication of standard dynamic programming techniques. To fill this gap, we propose a restless bandit formulation of the problem and propose an index policy for its solution. The problem is formally introduced in Section \ref{sec:Formulation}. After the so-called indexability property is established in Section \ref{sec:Indices} for the general case, closed form indices for specific families of the problem are provided in Section \ref{sec:Formulas}. In Section \ref{sec:Numerical}, the performance of the resulting index policy is tested in a series of numerical examples. Finally, Section \ref{sec:Conc} concludes the article.}

\section{\cm{The Multi-Machine Maintenance Problem with Imperfect Interventions}}\label{sec:Formulation}

\begin{figure}[!h]
\centering
\fbox{\includegraphics[trim={0 1cm 0 1cm}, width=8cm]{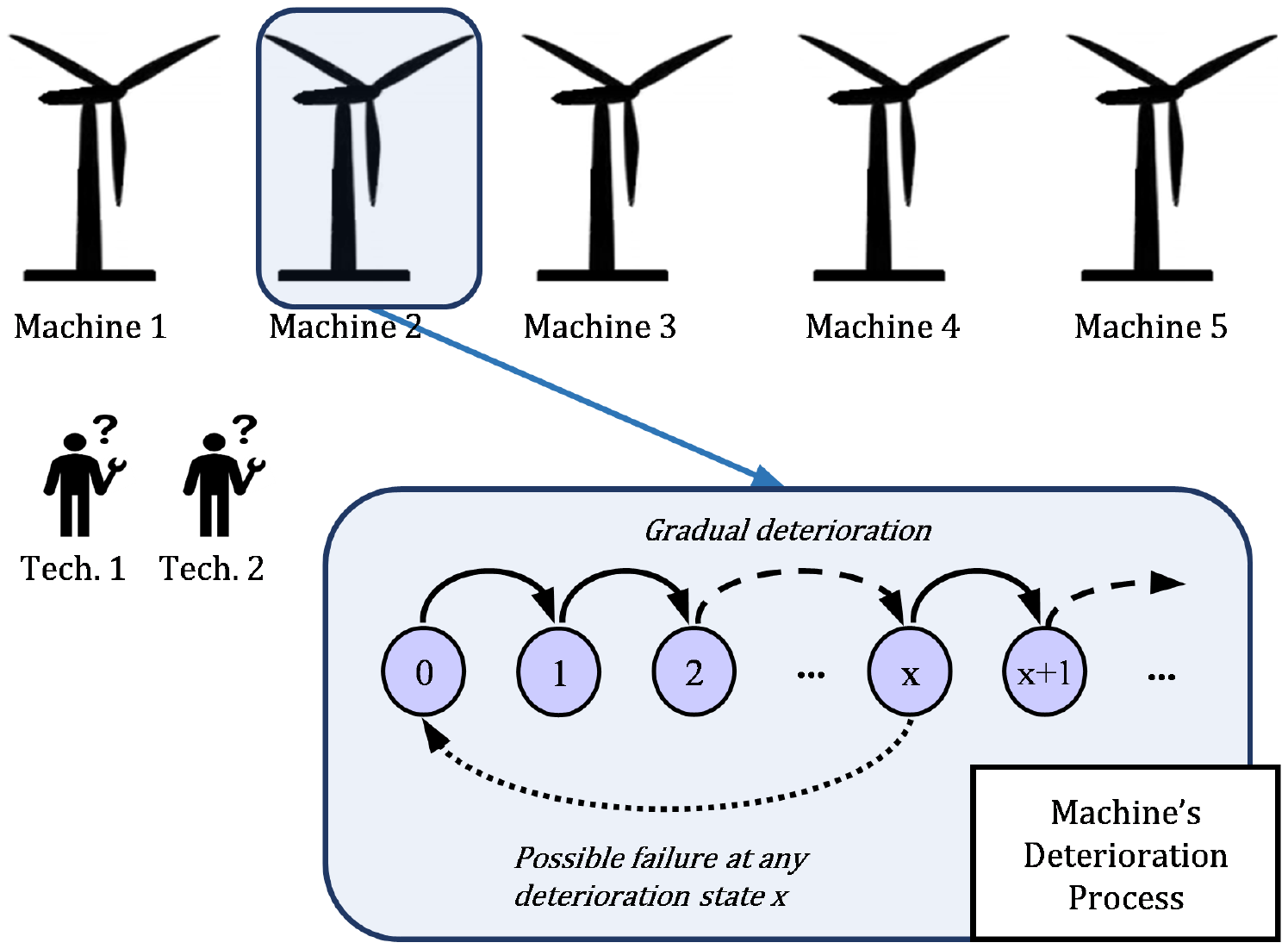}}
\caption{\cm{Illustrating example: five machines and two technicians.}}\label{SystemPic}
\end{figure}

\cm{Consider the problem of a manager that periodically must schedule preventive maintenance interventions in his equipment (sketched in Figure \ref{SystemPic}). Assume that the number of machines is larger than the number of repairmen (or teams) and, therefore, not all machines can be intervened simultaneously. Therefore, at each decision epoch the manager must select a collection of machines to be intervened during the corresponding period. This decision is typically based on information on the current state of the equipment obtained, for example, through condition monitoring systems.}

\cm{Machines deteriorate with use over a discrete space representing the state of wear of the equipment. It is assumed that machines operate and deteriorate independently from each other. Degradation increases both operation costs and failure probability.  Upon failure, the machine is either replaced by a new one or subject to a thorough repair that takes it back to a pristine state. Preventive and corrective maintenance tasks are assumed to be conducted by different dedicated units. }

\cm{Notwithstanding preventive maintenance interventions are aimed at taking the machine to a \textit{good-as-new} state, hidden faults not detected during the intervention, human error, faulty parts, or other external factors, can cause an intervention to be imperfect. In such cases the machine may be left in a better condition than the current one but not \textit{as-good-as-new}.  Preventive maintenance interventions are assumed immediate, in the sense that they are completed before the next decision epoch (when the machine will become available again). Maintenance costs are non-decreasing in the state of wear of the machine.}

\subsection{\cm{Mathematical Formulation}}

In this section, we \cm{present} the mathematical formulation of the problem described above. Consider that the manager seeks to schedule maintenance interventions in a collection $M$ of independently deteriorating machines by a set $R$ of repairmen. At each decision epoch $t \in \N$ the manager must choose a subset \cm{$Q\prn{t} \subseteq M$} of machines for intervention. Notice that the choice of $Q$ may allow the manager to leave some repairmen idle if necessary, i.e. \cm{$\abs{Q\prn{t}}\leq \abs{R}$}. As discussed in \cite{Ruiz}, the manager's problem can be modelled a discrete-time infinite horizon Markov Decision Problem, where each machine is modelled as a two action (operation/intervention) Markov decision chain that evolves over a discrete state space. \cm{States represent different (increasing) deterioration levels \cm{(see, for example, \citep{GlaRuKi,KurKha2010})}, with $0$ representing the \textit{as-good-as-new} state.} During the operation phase, the machine deteriorates, operation costs increase and the failure probability becomes larger. The intervention, which is conducted at a cost, aims at returning the machine to its \cm{ \textit{as-good-as-new}} state. However, due to unforeseeable reasons, the intervention may not be perfect, with the machine ending up at a more advanced state of wear than intended. The objective of the manager is minimising the discounted operation/intervention cost of the system over an infinite horizon.

The manager's problem can be characterised by $\cur{\beta,M,R,\xi,\mathcal{A},\mathcal{P},\mathcal{C}}$; where $\beta$ is a discount factor; $M$ and $R$ represent the sets of machines and repairmen, respectively; \cm{set $\xi=\cartes{S}{\abs{M}}{}$ denotes the system's state space, with $S_m$ representing the state space of machine $m=1,\ldots, \abs{M}$. The system's state at any decision epoch $t$ is given by $\mb{X}\prn{t}=\cur{X_m\prn{t}\in S_m, m=1,\ldots,\abs{M}}$. Set $\mathcal{A}$ represents the collection of all admissible actions $\mb{a}=\prn{a_1,\ldots,a_{\abs{M}}}$ at any given state $\mb{X}\prn{t}$, where $a_m\in \cur{0,1}$, for all $m=1,\ldots, \abs{M}$. Notice that $a_m$ represents the action to be taken at machine $m$, with value $0$ for operation and $1$ for intervention. Moreover, any admissible action $\mb{a}$ must satisfy the condition $\sum_{m=1}^{\abs{M}} a_m\leq \abs{R}$.} $\mathcal{P}$ is the set of all transition probability matrices $\mb{P}^{\mb{a}}$; where $\mb{P}^{\mb{a}}\prn{\mb{X},\mb{X'}}$ represents the probability of a transition from state $\mb{X}$ to state $\mb{X'}$  under action $\mb{a} \in \mathcal{A}$, for all $\mb{X},\mb{X'} \in \xi$. Finally, $\mathcal{C}$ is the collection of all possible vectors $\mb{C}^{\mb{a}}\prn{\mb{X}}$, representing the costs incurred from taking action $\mb{a} \in \mathcal{A}$ when the system is in state $\mb{X} \in \xi$. Notice that $P^a_m$ and $C^a_m$ represent the sets of transition and cost matrices, respectively, for any given machine $m$. 

Given this framework, the objective of the manager is to choose the stationary, deterministic policy $\pi$ that minimises the expected discounted operation/intervention cost over an infinite horizon, namely
\begin{align}\label{eq:MDP}
V^*=\min_{\pi}\cur{E_{\pi}\bra{\sum_{t=0}^{\infty} \beta^t \mb{C}^{\mb{a}\prn{t}} \prn{\mb{X}\prn{t}}}} 
\end{align}
Problem \eqref{eq:MDP} is a typical example of the so-called curse of dimensionality, which hinders the application of standard dynamic programming techniques for its solution. However, because of the independent evolution of the machines in the system, the maintenance scheduling problem can be addressed as a \textit{restless bandit problem} \cite{GlaRuKi}, and Whittle's extension \cite{Whit3} to the standard index theory can be deployed in order to find a good (near-optimal) solution to the operation/intervention problem. According to this theory if -after establishing the indexability of the problem- an index is calibrated for each machine/state pair in the system, then the policy that prescribes intervention in the subset of machines with larger indices is asymptotically optimal \cite{Whit3,WebWei}. \cm{In the following lines, we present the restless bandit formulation of our problem. The indexability of the problem and the structure of indices are discussed in Section \ref{sec:Indices}.}

As indexability and indices are properties of individual machines, in what follows we isolate an individual machine from the MDP problem described above and drop the machine indicator $m$. This machine is characterised by \cm{ $\cur{\beta,S,a,P^a,C^a}$}. In order to establish indexability and computing the corresponding indices, we develop a parameterised version of the problem by introducing a \textit{charge for intervention} $W$, which generates a family of cost-discounted MDPs for the machine.  The actions available when the machine is at state $X\prn{t}$ are given by $a\in\cur{0,1}$. If action $a=0$ (operation) is taken at decision time $t$, the machine performs a transition to some state $X'$ with probability $P^0\prn{X,X'}$; if, otherwise, action $a=1$ (intervention) is taken, then the system's state goes to some state $X''$ with probability $P^1\prn{X,X''}$. Under action $a=0$ an operation cost $C^0$, represented by $K\prn{X,X'}$, is incurred; likewise intervention implies a cost $C^1$, given by $C\prn{X}+W$. Notice that $K:S^2\rightarrow \R$ and $C:S\rightarrow \R$ are bounded continuous functions. Given these elements, the optimality equations for the $W$-charge problem evaluated at state $X\prn{t}=x$ can be expressed as

\begin{multline}\label{eq:ValF}
 V\prn{x,W}=\min\left\{K\prn{x}+\beta \sum_{x' \in S}P^0\prn{x,x'}V\prn{x',W},\right. \\
 \left. C\prn{x}+W+\beta \sum_{x'' \in S}P^1\prn{x,x''}V\prn{x'',W}\right\}, \quad x \in S
\end{multline}
where $K\prn{x}=\sum_{x' \in S} K\prn{x,x'}P^0\prn{x,x'}$ represents the expected cost incurred when operation is conducted at time $t$. This expectation is taken over all possible arrival states when starting from $X\prn{t}=x$.

\cm{With these elements we can now introduce the notion of indexability for the $W$-charge problem. Use $\Pi\prn{W}$ for the subset of $S$ for which the operation is optimal under intervention charge $W$; i.e. the first term of the expression between brackets in equation \eqref{eq:ValF} is larger or equal than the second term.  Based on Whittle's definition of indexability \cite{Whit3} we introduce a definition suitable for the characteristics of our model:}

\begin{definition}[\cm{\textbf{Indexability}}]\label{Indexability}
  Machine $\cur{\beta,S,a,P^a,C^a}$ is indexable if $\Pi\prn{W}$ is increasing in W, namely,
\[W_1\leq W_2 \Rightarrow \Pi\prn{W_1}\subseteq \Pi\prn{W_2}\]
\end{definition}
In words, a machine is indexable if, as the intervention charge increases, so does the collection of states for which operation is optimal. Moreover, the restless bandit formulation of the operation/intervention problem is indexable if each of its constituent bandits in indexable.

\begin{definition}[\cm{\textbf{Whittle Indices}}] \label{Whittle}
 If machine $\cur{\beta,S,a,P^a,C^a}$ is indexable, then its index $W:S\rightarrow \R$ is given by
\[W\prn{x}=\inf\cur{W:x\in \Pi\prn{W}}, \qquad x \in S\,.\]
\end{definition}
Namely, the index for a particular state $x$ is given by the minimal value of the intervention charge, $W$, for which it is optimal to operate according to equation \eqref{eq:ValF}. We refer to index $W\prn{x}$ as the $W$-index for state $x$.  Notice that the boundedness of the costs guarantees that the Whittle index will also be bounded. Intuitively, the index $W\prn{x}$ represents a \textit{fair charge for intervention} in state $x$, in the sense that it renders both actions, operation and intervention, optimal in the $W$-charge problem. 
\begin{definition}[\cm{\textbf{Index Policy}}]
Consider an indexable operation/intervention problem with $W$-index $W_m$ for machine $m, \; 1\leq m \leq M$ given by definition \ref{Indexability}. The \textit{index heuristic} prescribes, at each time $t\in \N$, to intervene in the $\abs{R}$ machines with largest index $W_m\prn{X\prn{t}}$, and to operate the remaining $\abs{M}-\abs{R}$ machines.
\end{definition}
\cm{Roughly speaking, an index heuristic consists of calibrating a set of state-dependent indices for each machine in the system, and -at each decision epoch- allocating the effort to the subset of machines with largest indices (further discussion on the index policy is provided at the end of Section \ref{sec:Indices} and sketched in Figure \ref{fig:IndPol}). Empirical studies testify the outstanding performance of the index policies in different areas of application \cite{AbIy2016,KG02,GlaMitAns}.}

\vspace{6pt}

\cm{\noindent \textbf{Comment}}

\cm{It is important noticing that the index policy is flexible enough to allow modelling situations that may arise in real life problems. For example, as the index is built independently for each machine, without making any assumption on the number of machines in the system, arrivals of new machines or temporary removals of damaged equipment can be easily incorporated in the model. In those cases, the set $M$ is simply extended or reduced correspondingly. Likewise, the number of repairmen available at each decision epoch is not required to be the same. For example, if a number $r$ of technicians is not available at a particular epoch, the decision can simply be made selecting the $\abs{R}-r$ machines with larger index.}

\vspace{6pt}

Once the model has been described and the main concepts of indices and indexability introduced, in the next section we develop a collection of suitable indices for the operation/intervention problem and establish a general form of indexability.

\section{Indices and Indexability Analysis}\label{sec:Indices}

In order to obtain the $W$-indices for the restless bandit formulation of the operation/intervention problem, in the following sections we analyse two possible courses of action that can be taken whenever a machine is in state $x \in S$ at time $t$. \cm{It will be argued that the $W$-index for state $x$ is given by the value of the activity charge $W$ for which both policies render identical costs.}

\subsection{Cost of immediate intervention: the $\B\prn{x,W}$-policy}\label{sec:B-Pol}

We start by considering a set up where, independently of the initial state of the machine and after following an optimal (cost minimising) policy, whenever the machine arrives to state $X\prn{t}=x$, it is optimal to perform an intervention. This is summarised in the following assumption.

\begin{assumption} 
 If intervention is optimal at state $X\prn{t}=x$ when operating from $X\prn{0}=0$, then it is optimal when $X\prn{0}=y$, for any $\; 0 < y \leq x$.
\end{assumption}

\begin{figure}[!ht]
\centering
\fbox{\includegraphics[width=8cm]{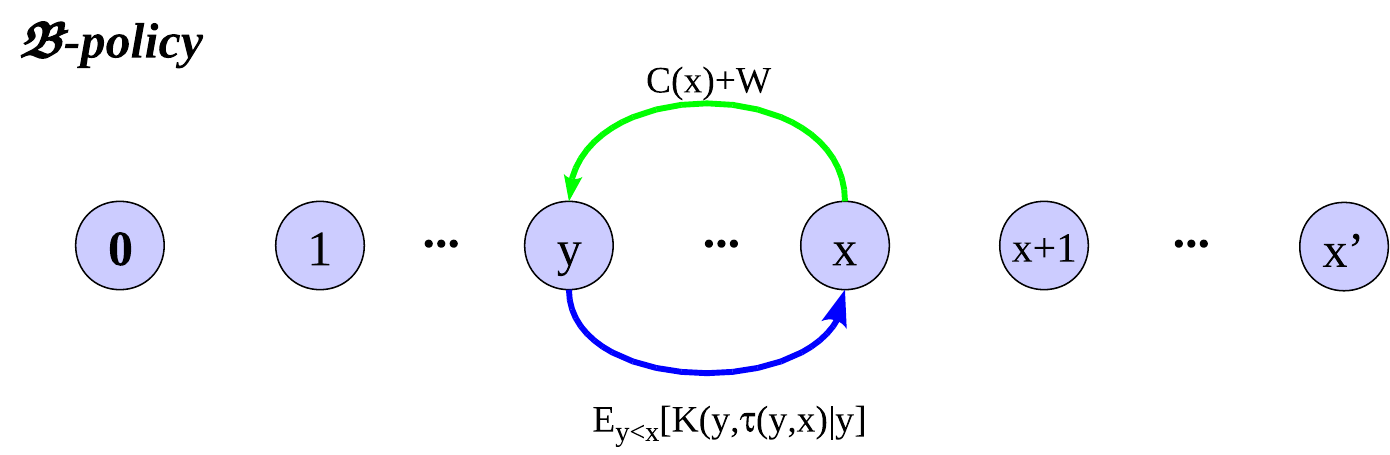}}
\caption{Machine's evolution under $\B$-policy}\label{fig:B-Pol}
\end{figure}

It is convenient here to recall that, notwithstanding maintenance interventions are aimed at taking the machine back to an \textit{as-good-as-new} state, unforeseeable events may affect the outcome of the intervention and the machine may end-up in a more advanced state of wear than expected. The evolution of the machine after intervention is illustrated in Figure \ref{fig:B-Pol} and can be summarized in steps \eqref{s1} to \eqref{s3}. 

\begin{enumerate}[i)]
 \item \label{s1} Intervention has taken the machine to certain state $y<x$ with probability $P^1\prn{x,y}$. The machine is operated optimally until the first entry, after $y$, back to state $x$. This takes a time represented by $\tau\prn{y,x}$. The expected discounted cost incurred during a transition from $y$ to $x$ is given by
\[K\prn{y,\tau\prn{y,x}}=E\bra{\sum_{t=0}^{\tau\prn{y,x}-1}\beta^{t}K\prn{X\prn{t}}|y}\]
where the conditional $|y$ is a notational shorthand for $|X\prn{0}=y$.

Therefore, the expected cost of returning to state $x$ after a maintenance intervention becomes
\[E_{y<x|x}\bra{K\prn{y,\tau\prn{y,x}}} \,.\]
where the expectation is taken over all possible values of $y$ such that $y<x$, for all $x,y\in S$.

 \item Once the machine has returned to state $x$, a new maintenance intervention is performed with expected cost given by
\[E_{y<x|x}\bra{E\beta^{\tau\prn{y,x}}\prn{C\prn{x}+W}|y}\]

\item \label{s3} The intervention takes the system to certain state $y<x$, from which the policy described above is repeated indefinitely.
\end{enumerate}

The expected discounted cost of such policy, incurred over the infinite horizon, satisfies the equation:
\begin{multline*}
E_{y<x|x}B\prn{x,W}=E_{y<x|x}\bra{K\prn{y,\tau\prn{y,x}}}\\ +E_{y<x|x}\bra{E\beta^{\tau\prn{y,x}}\prn{C\prn{x}+W+E_{y<x|x}B\prn{x,W}}|y} 
\end{multline*}

or, equivalently 
\begin{align}\label{B}
E_{y<x|x}B\prn{x,W}+W=\frac{W+E_{y<x|x}\bra{K\prn{y,\tau\prn{y,x}}} +E_{y<x|x}\bra{E\beta^{\tau\prn{y,x}}C\prn{x}|y} }{1-E_{y<x|x}\bra{E\beta^{\tau\prn{y,x}}|y}}
\end{align}

We finally define:
\begin{align}\label{def:B}
\B\prn{x,W}=E_{y<x|x}B\prn{x,W}+W\,.
\end{align}

\begin{proposition}\label{ass:W_incr}
 Quantity $\B\prn{x,W}$ is continuous and strictly increasing in $W$. 
 
 \begin{proof}
  The continuity of $\B$ follows directly from the definition of $B\prn{x,W}$ and the continuity of $K\prn{X}$ and $C\prn{X}$. That $\B$ is increasing in $W$ follows straightforwardly from \eqref{B}.
 \end{proof}
\end{proposition}

\subsection{Cost of postponing intervention: the $G\prn{x}$-policy}

\begin{figure}[!ht]
\centering
\fbox{\includegraphics[width=8cm]{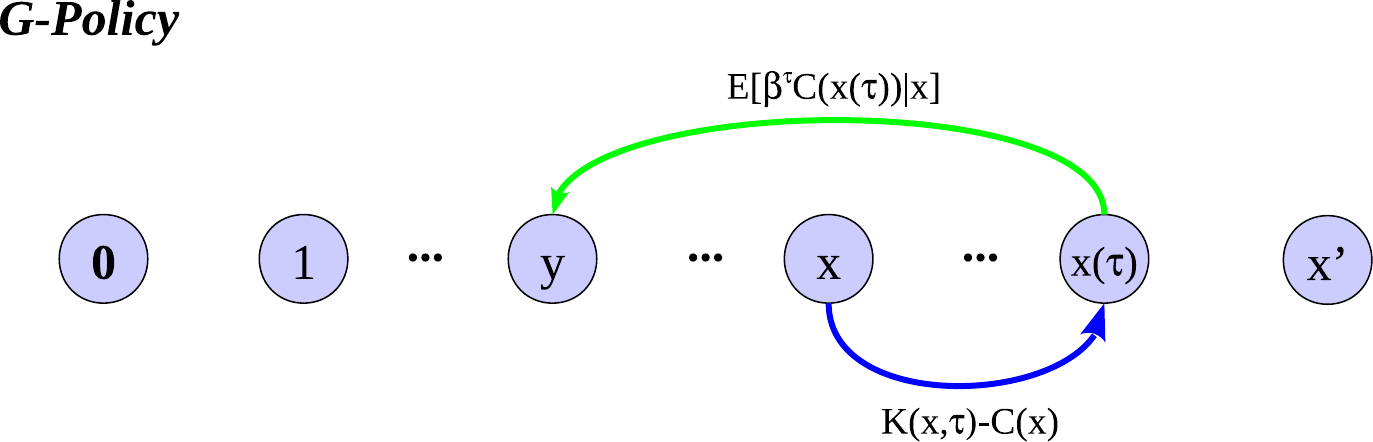}}
\caption{Machine's evolution under G-policy}\label{fig:G-Pol}
\end{figure}

Consider a set up where, starting from state $X\prn{0}=x$, the machine is operated for $\tau$ additional periods, where $\tau\geq 1$ almost surely, after which a maintenance intervention is carried out. The evolution of the machine under this scenario is depicted in Figure \ref{fig:G-Pol}. The expected incremental cost incurred from $t=0$ until the machine returns to state $x$ is given by
\[K\prn{x,\tau}+E\bra{\beta^{\tau}C\prn{X\prn{\tau}}|x}-C\prn{x}\]

This framework allows us to introduce a form of Gittins index \cite{Gittins, GitJon} appropriate for our analysis:

\begin{definition}\label{def:Git}
 The Gittins index for operation $G:S\rightarrow \R$ in state $x\in S$ is given by
\begin{align}\label{Git}
 G\prn{x}=\inf_{\tau}\cur{\frac{K\prn{x,\tau}+E\bra{\beta^{\tau}C\prn{X\prn{\tau}}|x}-C\prn{x}}{1-E\bra{\beta^{\tau}|x}}}
\end{align}
where the infimum in \eqref{Git} is taken over all positive-valued stopping times on the machine state process evolving from $x$ under operation.
\end{definition}

\cm{This index can be interpreted as the optimal operation/intervention cost rate that can be incurred by a machine when its initial state is $X\prn{0}=x$. A thorough discussion on the characterisation of the set of stationary stopping times achieving the infimum in \eqref{Git} can be found in \cite{GlaRuKi,GitGlaWe2011}.}

Before proceeding with our main indexability result, we need to introduce the following assumption:

\begin{assumption}
 Whenever $X\prn{0}=x$ and operation is conducted during $\tau$ additional periods, then  $P^1\prn{X\prn{\tau},y}\approx0$ for all $x \leq y \leq X\prn{\tau}$; i.e. we assume that a maintenance intervention, taken after an additional operation lap, will take the system to a state $y< x$ almost surely.
\end{assumption}
 
 \subsection{W-Indices}

\cm{The following framework will help developing a general expression for the $W$-indices. Consider a policy that, given an initial state $X\prn{0}=x$ and intervention time $\tau$: operates the equipment at times $t=0,\ldots,\tau-1$; carries out preventive maintenance at time $\tau$; and follows an optimal operation/intervention policy at any other subsequent time $t>\tau$. This policy is optimal if and only if there exists some stationary stopping time $\tau>0$, on the machine operation process $\cur{X\prn{t},t\geq0}$, such that the policy's total expected discounted cost is no greater than that of a policy that chooses intervention at $t=0$ and behaves optimally from there on.}

The decision faced by the manager is whether to intervene at time $t=0$, taking the machine to certain (unknown) state $y<x$, and then to follow the cost minimising policy described in points \eqref{s1}-\eqref{s3} incurring a minimum expected cost given by
\[W+C\prn{x}+E_{y<x|x}B\prn{x,W}\,;\]
or to operate the machine for some additional time $\tau$, at which point intervention is conducted, operating optimally thereafter. The expected cost of such policy is given by
\begin{align*}
 K\prn{x,\tau}+E\bra{\beta^{\tau}\prn{C\prn{X\prn{\tau}}+W}|x}+E\bra{\beta^{\tau}|x}E_{y<x|x}B\prn{x,W}\,.
\end{align*}
Operation will be an optimal choice in $x$ if and only if there exists an stationary stopping time, $\tau>0$, such that
\begin{multline*}
  K\prn{x,\tau}+E\bra{\beta^{\tau}\prn{C\prn{X\prn{\tau}}+W}|x}+E\bra{\beta^{\tau}|x}E_{y<x|x}B\prn{x,W} \\ \leq W+C\prn{x}+E_{y<x|x}B\prn{x,W}
\end{multline*}
i.e. $\exists \;\tau >0$ s.t.
\begin{multline*}
  K\prn{x,\tau}+E\bra{\beta^{\tau}C\prn{X\prn{\tau}}|x}+E\bra{\beta^{\tau}|x}\prn{E_{y<x|x}B\prn{x,W}+W}-C\prn{x} \\ \leq \B\prn{x,W}
\end{multline*}
i.e. $\exists \;\tau >0$ s.t.
\begin{align}\label{Equil}
  \frac{K\prn{x,\tau}+E\bra{\beta^{\tau}C\prn{X\prn{\tau}}|x}-C\prn{x}}{1-E\bra{\beta^{\tau}|x}} \leq \B\prn{x,W}
\end{align}

\begin{figure}[!ht]
\centering
\fbox{\includegraphics[width=8cm]{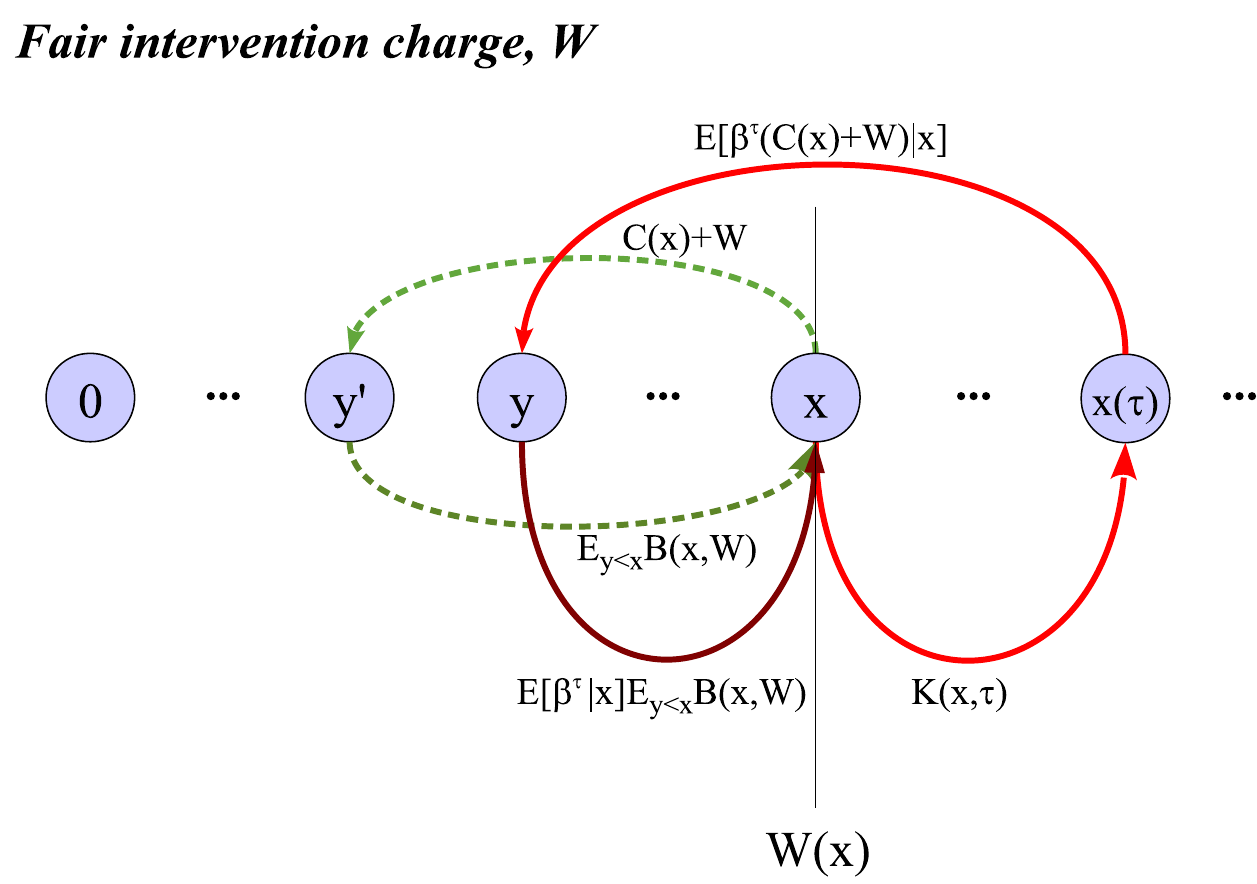}}
\caption{Determination of $W$-index from policies $\B\prn{x,W}$ (dashed) and $G\prn{x}$ (solid).} \label{fig:W-Ind}
\end{figure}

Notice that, if we take the \textit{infimum} over all positive stopping times $\tau$, then the left hand side term in \eqref{Equil} is the \textit{Gittins Index} \eqref{def:Git} for state $x \in S$. Moreover, from Definition \ref{def:Git} and from the fact that the infimum in \eqref{Git} is always achieved, the left hand side of expression \eqref{Equil} is met precisely when
\[G\prn{x} \leq \B\prn{x,W}.\]
As it will be stated in Theorem \ref{the:ind}, it follows naturally from the discussion above that the $W$-index for state $x$ is given by the $W\prn{x}$ solution to the equation $G\prn{x} = \B\prn{x,W}$. The interaction between $\B\prn{x,W}$ and $G\prn{x}$ is illustrated in Figure \ref{fig:W-Ind}.

In order to establish our indexability result, and to derive an expression for the $W$-index, we first need to define set $\Pi\prn{W}$ for the $W$-charge problem:

\begin{definition}
\textbf{Operation Set} \\
We define the operation set
\begin{align}\label{PasSet}
\Pi\prn{W}=\cur{x \in S: G\prn{x} \leq \B\prn{x,W}}
\end{align}
as the set of states $x\in S$ for which operation is optimal in the $W$-charge problem, given an intervention charge $W$. 
\end{definition}
Following Whittle's discussion in \cite{Whit3} and according to Definition \ref{Indexability}, the machine $\cur{\beta,S,a,P^a,C^a}$  --and the associated operation/intervention problem-, will be indexable if $\Pi\prn{W}$ is increasing in $W$, namely, the set of machine states where it is optimal to operate, for a given intervention charge $W$, is increasing in $W$.

\begin{theorem}\label{the:ind}
 \textbf{Indexability and Indices}

\begin{enumerate}[a)]
 \item Machine $\cur{\beta,S,a,P^a,C^a}$ is indexable.
 \item The W-index for state $x$, denoted by $W\prn{x}$, is the unique $W$ solution to the equation
\[G\prn{x} = \B\prn{x,W} \]
 \item The orderings of members of $S$ determined by the $W$-index and the Gittins index, $G$, coincide.
\end{enumerate}

\begin{proof}
By Proposition \ref{ass:W_incr}, $\B\prn{x,W}$ is strictly increasing in $W$. It then follows from \eqref{PasSet} that $\Pi\prn{W}$ is increasing, and indexability follows immediately from Definition \ref{Indexability}. It also holds from the continuity of $\B\prn{x,W}$ and Definition \ref{Whittle} that the W-index for state $x$, namely
\[W\prn{x}=\inf\cur{W:x \in \Pi\prn{x}}\]
satisfies the equation
\begin{align} \label{eq:W}
 G\prn{x} = \B\prn{x,W\prn{x}}
\end{align}
By the strictly increasing nature of $\B\prn{x,W}$, equation \eqref{eq:W} specifies $W\prn{x}$ uniquely. This establish parts (a) and (b) of the theorem. Part (c) follows simply from the fact that $W\prn{x}$ is strictly increasing in $G\prn{x}$.
\end{proof}
\end{theorem}

\cm{A general interpretation of the index is that $W^m\prn{x}$ represents a fair charge for intervening machine $m$ when its current stat is $X\prn{t}=x$. A more intuitive interpretation of $W\prn{x}$ will be provided in the discussion around equation \eqref{eq:Whit} in the following Section. }

\begin{figure}[!h]
\centering
\fbox{\includegraphics[trim={1.6cm 5.5cm 1cm 1cm}, width=8cm]{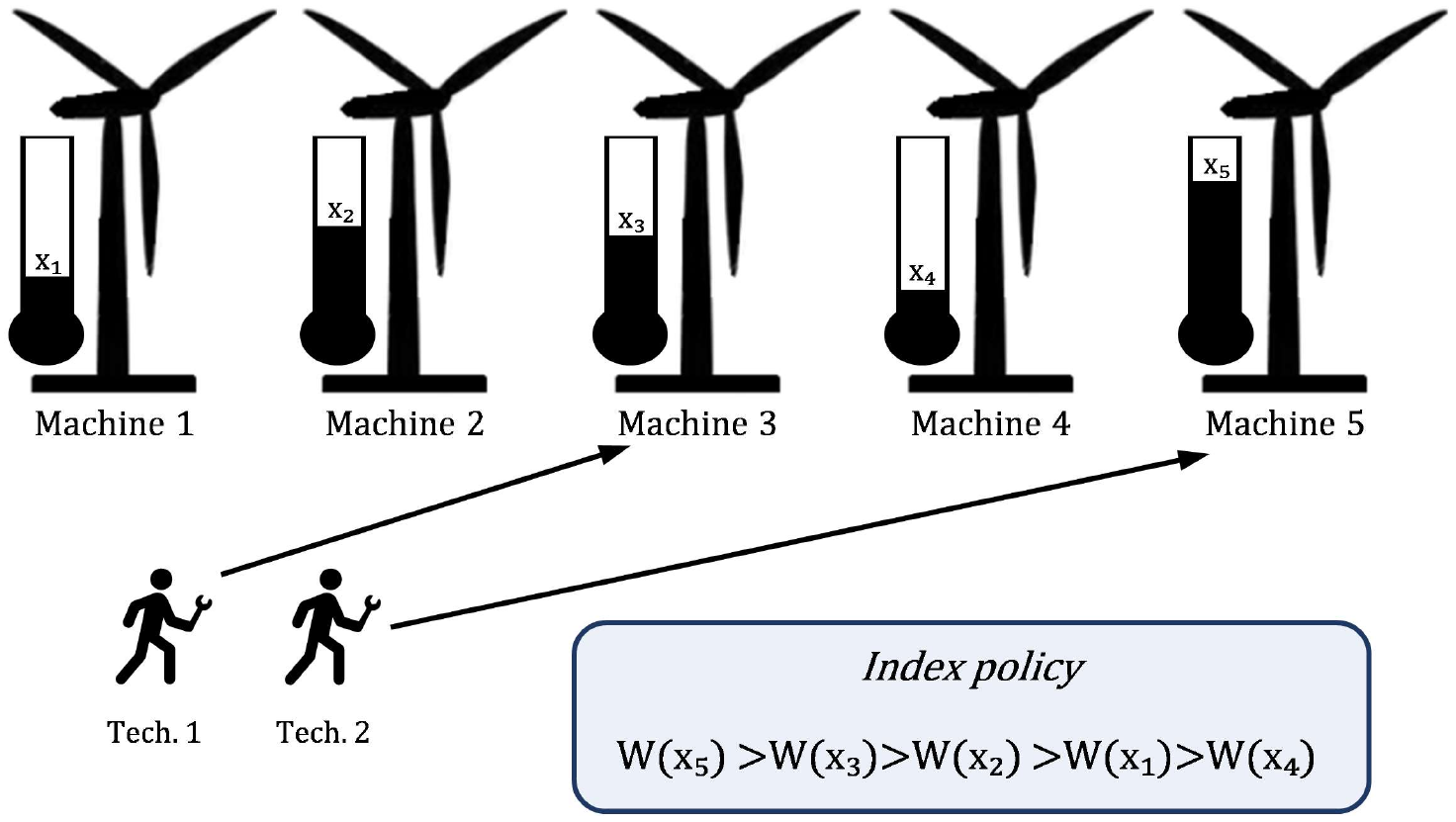}}
\caption{\cm{Illustrating example: index policy.}}\label{fig:IndPol}
\end{figure}

\cm{Once the indices have been computed, the index policy prescribes carrying out preventive maintenance in those machines whose current state returns larger values of the $W$-index. Notice that the indices are built considering not only the current state of wear of the machine, but also the machine's specific characteristics and its possible paths of future evolution; this implies that ordering the machines by their index value will not necessarily return the same sequence than the one obtained when the ordering criteria is the current state. This is sketched in Figure \ref{fig:IndPol}, where notwithstanding machine 2 shows a higher state of wear than machine 3, its index is smaller and, consequently, the index policy prescribes intervention in machines 5 and 3.}

Subsequent analysis will focus on developing the $W$-indices under specific problem's characteristics.

\section{\cm{Closed-form expressions for the $W$-indices}}\label{sec:Formulas}

Once the main notions of indexability and indices have been developed, we now proceed to introduce some particular elements of the machine maintenance problem with imperfect interventions that will be useful for developing closed-form expressions for the corresponding indices.

\cm{We assume that between interventions the machine is subject to gradual deterioration, sporadic failures, and non-decreasing operation costs. Preventive interventions are comprehensive and are aimed at taking the machine back to its \textit{as-good-as-new} state. However, due to unforeseeable factors, the machine may end up in a more advanced state of wear than the one pursued.  Maintenance interventions are conducted by a group of repairmen (or teams) whose number is smaller than the number of machines and, consequently, not all machines can be intervened at each decision epoch. Maintenance costs include foregone productivity, intervention related expenses and parts. Any failure is followed by thorough repair or immediate replacement of the machine, incurring a large cost.}

To help characterising our model, we introduce the following assumptions:
\begin{assumption}
 The state space, $S$, is the natural numbers ($\N$), with $0$ representing the so-called \textit{as-good-as-new} state.
\end{assumption}
\begin{assumption}\label{ass:Probs}
 Evolution under operation is right skip free, namely,
\[P^0\prn{x,x}+P^0\prn{x,x+1}+P^0\prn{x,0}=1\]
where $P\prn{x,0}$ represents the probability of a failure. Moreover, as machines are deteriorating at a non-decreasing rate, it must hold that $P^0\prn{x,x+1}\leq P^0\prn{x+1,x+2}$ and $P\prn{x,0}\leq P\prn{x+1,0}, \, x \in S$.
\end{assumption}

\cm{The right-skip free condition simply guarantees a gradual degradation of the machine. Notice that when the machine is at its \textit{as-good-as-new} state there are no failures and, consequently, $P^0\prn{0,0}+P^0\prn{0,1}=1$, where $P^0\prn{0,0}$ simply represents the probability of a new sojourn in state $0$.}

\begin{assumption}\label{ass:Interv}
 Evolution under intervention is given by
 \[\sum_{y<x}P^1\prn{x,y}=1, \quad x,y \in S\,; \]
moreover, it holds that $P^1\prn{x,y} > P^1\prn{x,y+1}$ for all $y<x,\, x,y \in S$.
\end{assumption}

The comment in Assumption \ref{ass:Interv} represents the intuitive fact that, when intervention is conducted at given state, large errors are less likely than smaller ones.
\begin{assumption}\label{ass:IncGit}
The index for operation $G:\N \rightarrow \R$ is (strictly) increasing.
\end{assumption}

\cm{Suppose that $X\prn{0}=x$ and that the machine state evolves under the passive action. We use $\tau\prn{x,x'}$ for the time to the first entry into state $x'$ when the system starts at $x$. Notice that because of the \textit{right skip free} assumption, $\tau\prn{x,x'}<\tau\prn{x,x'+1}$ for all $x<x'$, implying that the time required for a machine departing from state $x$ for visiting state $x'$, is shorter than the time required for visiting $x'+1$.}

The stopping time achieving the infimum in \eqref{Git} can be written as
\begin{align*}
 \tau^*=\min\cur{t: t\geq1, G\prn{X\prn{t}}\geq G\prn{x}}
\end{align*}
additionally, because of Assumption \ref{ass:IncGit}, it holds that
\begin{align*}
 \tau^*_x=\min\cur{\tau\prn{x,x},\tau\prn{x,x+1}}
\end{align*}
represents the required time until the first visit to $x$ or $x+1$ when operation is conducted from state  $x$.

Given that $\tau^*_x$ achieves the infimum in \eqref{Git}\footnote{For a thorough discussion of this property please refer to Lemma 4.1 in \cite{Ruiz}.} the index $G\prn{x}$ can be rewritten as
\begin{align}\label{eq:Git}
 G\prn{x}=\frac{K\prn{x,\tau^*_x}+E\bra{\beta^{\tau^*_x}C\prn{X\prn{\tau^*_x}}|x}-C\prn{x}}{1-E\bra{\beta^{\tau^*_x}|x}}
\end{align}

With this result and using the definition of $\B\prn{x,W}$ in \eqref{def:B} we can write the following definition of the W-Index:

\begin{theorem}{\textbf{W-Indices}}\label{Theo:WhitMon}
 
The W-index for machine $\cur{\beta,S,a,P^a,C^a}$ in state $x$, in the operation/intervention problem with imperfect maintenance, is given by 
\begin{multline}\label{eq:Whit}
W\prn{x}=G\prn{x}\prn{1- E_{y<x|x}\bra{\beta^{\tau\prn{y,x}}}|y} -E_{y<x|x}\bra{K\prn{y,\tau\prn{y,x}}} \\ -E_{y<x|x}\bra{E\beta^{\tau\prn{y,x}}C\prn{x}|y}
\end{multline}
for all $x \in \N^+$, where $G\prn{x}$ is given by \eqref{eq:Git}. $W\prn{x}$ is increasing in $x$.
 
\begin{proof}
 By Theorem \ref{the:ind}, $W\prn{x}$ is the unique solution to $G\prn{x}=\B\prn{x,W}$. Solving 
\begin{align*}
 \B\prn{x,W}=\frac{W+E_{y<x|x}\bra{K\prn{y,\tau\prn{y,x}}} +E_{y<x|x}\bra{E\beta^{\tau\prn{y,x}}C\prn{x}|y} }{1-E_{y<x|x}\bra{E\beta^{\tau\prn{y,x}}|y}}=G\prn{x}
\end{align*}
for $W$ we get \eqref{eq:Whit}. Given that  $W\prn{x}$ is strictly increasing in $G\prn{x}$, and from Assumption \ref{ass:IncGit}, it holds that $W\prn{x}$ is increasing in $x$.
\end{proof}
\end{theorem}

\cm{The first term in the right hand side of equation \eqref{eq:Whit} represents the expected discounted cost of a policy that, upon arrival to state $x$, operates for an additional period of duration $\tau$ and then conducts a maintenance intervention; the other two terms represent the expected discounted cost of a policy that always intervenes in state $x$. The difference between them represents the cost savings incurred if intervention is taken immediately. This gives a natural interpretation of the index policy, which prescribes carrying out maintenance on those machines where intervention brings up larger cost savings.}

\vspace{6pt}

\cm{We now develop the explicit formulae for the index in the context of the working assumptions of the model outlined at the beginning of this section. We start by defining the following values:}
\begin{center}
\begin{tabular}{ccc}
$\delta\prn{x} = \frac{\beta P^0\prn{x,x+1}}{1-\beta P^0\prn{x,x}} \label{delta}$; & \qquad
$\kappa\prn{x}=\frac{K\prn{x}}{1-\beta P^0\prn{x,x}}$; & \qquad
$\gamma\prn{x}= \frac{\beta P^0\prn{x,0}}{1-\beta P^0\prn{x,x}}$
\end{tabular}
\end{center}
with $\gamma\prn{0}=0$.

To obtain an expression for $E\bra{\beta^{\tau^*_x}\left.\right|x}$ we first need to obtain $E\beta^{\tau\prn{0,x}}$. We start noticing that
\[
 E\beta^{\tau\prn{0,1}}=\beta P^0\prn{0,1}+\beta P^0\prn{0,0}E\beta^{\tau\prn{0,1}}=\delta\prn{0}
\]
and
\begin{align*}
E\beta^{\tau\prn{1,2}}&=\beta P^0\prn{1,2}+\beta P^0\prn{1,1}E\beta^{\tau\prn{1,2}}+\beta P^0\prn{1,0}\beta^{\tau\prn{0,1}}\beta^{\tau\prn{1,2}}\\
&=\frac{\beta P^0\prn{1,2}}{1-\beta P^0\prn{1,1}-\beta P^0\prn{1,0} E\beta^{\tau\prn{0,1}}} \\
&=\frac{\delta\prn{1}}{1-\gamma\prn{1}\delta\prn{0}}
 \end{align*}
consequently
\begin{align*}
 E\beta^{\tau\prn{0,2}}=E\beta^{\tau\prn{0,1}}E\beta^{\tau\prn{1,2}}=\frac{\delta\prn{0}\delta\prn{1}}{1-\gamma\prn{1}\delta\prn{0}}
\end{align*}
following the same reasoning, it can be easily verified that
\begin{align}\label{beta_0}
 E\beta^{\tau\prn{0,x}}=\frac{\prod_{y=0}^{x-1}\delta\prn{x}}{1-\sum_{y=0}^{x-1}\gamma\prn{y}\prod_{z=0}^{y-1}\delta\prn{z}}\,.
\end{align}
We now have that
\begin{align}\label{beta_star}
 E\bra{\beta^{\tau^*}|x}&=E\bra{\beta^{\tau_x\prn{x,x+1}}}=\beta \prn{1-P^0\prn{x,0}} + \beta P^0\prn{x,0} E\beta^{\tau\prn{0,x}}
\end{align}
and
\begin{multline}\label{C_x}
 E\bra{\beta^{\tau^*} C\prn{X\prn{\tau^*}}|x}=\beta P^0\prn{x,x} C\prn{x}+\beta P^0\prn{x,x+1}C\prn{x+1} \\ +\beta P^0\prn{x,0}E\beta^{\tau\prn{0,x}}C\prn{x}
\end{multline}
can easily be computed by direct substitution of \eqref{beta_0}.

In order to find $E_{y<x}\bra{\beta^{\tau\prn{y,x}}}$ we notice that
\begin{align}
 E\beta^{\tau\prn{x-1,x}}&=\beta P^0\prn{x-1,x}+\beta P^0\prn{x-1,x-1}E\beta^{\tau\prn{x-1,x}} \notag \\ &\hspace{2cm}+\beta P^0\prn{x-1,0}E\beta^{\tau\prn{0,x-1}}E\beta^{\tau\prn{x-1,x}} \notag \\
&=\frac{\delta\prn{x-1}}{1-\gamma\prn{x-1}E\beta^{\tau\prn{0,x-1}}} \,;\notag
\intertext{using \eqref{beta_0} and upon simplification we get}
&=\frac{ \delta\prn{x-1}\prn{1-\sum_{i=0}^{x-2}\gamma\prn{i}\prod_{z=0}^{i-1}\delta\prn{z}}}{1-\sum_{i=0}^{x-1}\gamma\prn{i}\prod_{z=0}^{i-1}\delta\prn{z}}  \label{beta_next}\,;
\end{align}
whereas for $x-2$ it holds that
\begin{align*}
E\beta^{\tau\prn{x-2,x}}=&\beta P^0\prn{x-2,x-1}E\beta^{\tau\prn{x-1,x}}+\beta P^0\prn{x-2,x-2}E\beta^{\tau\prn{x-2,x}} \\ 
&\hspace{4cm}+\beta P^0\prn{x-2,0} E\beta^{\tau\prn{0,x-2}}E\beta^{\tau\prn{x-2,x}}\\
=& \frac{\beta P^0\prn{x-2,x-1}E\beta^{\tau\prn{x-1,x}}}{1-\beta P^0\prn{x-2,x-2}-\beta P^0\prn{x-2,0}E\beta^{\tau\prn{0,x-2}}}\\
=& \frac{\delta\prn{x-2}E\beta^{\tau\prn{x-1,x}}}{1-\gamma\prn{x-2}E\beta^{\tau\prn{0,x-2}}}\\
=& \frac{\delta\prn{x-2}\delta\prn{x-1}\prn{1-\sum_{i=0}^{x-3}\gamma\prn{i}\prod_{z=0}^{i-1}\delta\prn{z}}}{1-\sum_{i=0}^{x-1}\gamma\prn{i}\prod_{z=0}^{i-1}\delta\prn{z}}\,.
\end{align*}
This expression can be easily generalised to
\begin{align}
E\beta^{\tau\prn{y,x}}= \frac{\displaystyle \prod_{h=y}^{x-1}\delta\prn{h}\prn{1-\sum_{i=0}^{y-1}\gamma\prn{i}\prod_{z=0}^{i-1}\delta\prn{z}}}{\displaystyle 1-\sum_{i=0}^{x-1}\gamma\prn{i}\prod_{z=0}^{i-1}\delta\prn{z}}\,;
\end{align}
and taking expectation over all possible values of $y<x$ we obtain
\begin{align}\label{expec_beta}
 E_{y<x}\bra{\beta^{\tau\prn{y,x}}}=\sum_{y=0}^{x-1}P^1\prn{x,y}\frac{\displaystyle \prod_{h=y}^{x-1}\delta\prn{h}\prn{1-\sum_{i=0}^{y-1}\gamma\prn{i}\prod_{z=0}^{i-1}\delta\prn{z}}}{\displaystyle 1-\sum_{i=0}^{x-1}\gamma\prn{i}\prod_{z=0}^{i-1}\delta\prn{z}} \,.
\end{align}

In order to obtain expressions for $K\prn{x,\tau^*_x}$ and $K\prn{y,\tau\prn{y,x}}$ we start with the fact that
\[K\prn{0,\tau\prn{0,1}}=K\prn{0}+\beta P^0\prn{0,0}K\prn{0,\tau\prn{0,1}}=\kappa\prn{0}\,.\]
We kindly ask the reader to recall that $K\prn{x}$ represents the cost incurred when operating the machine in state $x$ during one period; whereas $K\prn{x,\tau\prn{x,x'}}$ represents the cost incurred when the machine is in operation from state $x$ until a transition to state $x'$ occurs.

In order to compute
\[K\prn{0,\tau\prn{0,2}}=K\prn{0,\tau\prn{0,1}}+E\beta^{\tau\prn{0,1}}K\prn{1,\tau\prn{1,2}}\]
we notice that
\begin{align*}
 K\prn{1,\tau\prn{1,2}}&=K\prn{1}+\beta P^0\prn{1,1}K\prn{1,\tau\prn{1,2}}+\beta P^0\prn{1,0}K\prn{0,\tau\prn{0,2}}\\
&=K\prn{1}+\beta P^0\prn{1,1}K\prn{1,\tau\prn{1,2}}\\
& \hspace{2cm}+\beta P^0\prn{1,0}\prn{\kappa\prn{0}+E\beta^{\tau\prn{0,1}}K\prn{1,\tau\prn{1,2}}+B}\\
&=\frac{\kappa\prn{1}+\gamma\prn{1} \prn{\kappa\prn{0}+B}}{1-\gamma\prn{1}\delta\prn{0}}
\end{align*}
therefore
\begin{align*}
 K\prn{0,\tau\prn{0,2}}=\frac{\kappa\prn{0}+\delta\prn{0}\prn{\kappa\prn{1}+\gamma\prn{1}B}}{1-\gamma\prn{1}\delta\prn{0}}\,.
\end{align*}
Following the same reasoning, we conclude that
\begin{align}\label{K_0}
 K\prn{0,\tau\prn{0,x}}=\frac{\displaystyle\sum_{y=0}^{x-1}\prn{\kappa\prn{y}+\gamma\prn{y}B}\prod_{z=0}^{y-1}\delta\prn{z}} {1-\sum_{y=0}^{x-1}\gamma\prn{y}\prod_{z=0}^{y-1}\delta\prn{z}}\,;
\end{align}
and therefore
\begin{align}\label{K_star}
 K\prn{x,\tau^*_x}&=K\prn{x}P^0\prn{x,x}+k\prn{x}P^0\prn{x,x+1}\notag \\ & \hspace{3cm}+P^0\prn{x,0}\prn{k\prn{x}+K\prn{0,\tau\prn{0,x}}+B} \notag \\
&=K\prn{x}+P^0\prn{x,0}\prn{K\prn{0,\tau\prn{0,x}}+B}
\end{align}
can be obtained by direct substitution of \eqref{K_0}.

In order to compute $E_{y<x}K\prn{y,\tau\prn{y,x}}$ we first notice that for any $x \in S$, the expression
\begin{align}\label{K_next}
 K\prn{x,\tau\prn{x,x+1}}=\frac{\kappa\prn{x}+\gamma\prn{x}\prn{K\prn{0,\tau\prn{0,x}}+B}}{1-\gamma\prn{x}E\beta^{\tau\prn{0,x}}}
\end{align}
can be obtained by direct substitution of \eqref{beta_0} and \eqref{K_0}. Additionally, for $y=x-1,  \, x \in S$, $K\prn{x-1,\tau\prn{x-1,x}}$
can be computed directly using \eqref{K_next}, and for any other value $y<x-1$ the expression
\begin{align}
 K\prn{y,\tau\prn{y,x}}=K\prn{y,\tau\prn{y,y+1}}+E\beta^{\tau\prn{y,y+1}}K\prn{y+1,\tau\prn{y+1,x}}
\end{align}
can be obtained recursively using  \eqref{K_next} and \eqref{beta_next}. From here the computation of
\begin{align}\label{eq:Exp_K}
 E_{y<x}K\prn{y,\tau\prn{y,x}}=\sum_{y=0}^{x-1}P^1\prn{x,y} K\prn{y,\tau\prn{y,x}}
\end{align}
is straightforward.

Using expressions \eqref{beta_star},\eqref{C_x}, and \eqref{K_star} in \eqref{eq:Git} we can define the function $H: \N \rightarrow \R$ as:
 \begin{align*}
 H\prn{x}&=\left[\rule{0pt}{18pt} K\prn{x}+P^0\prn{x,0}\prn{K\prn{0,\tau\prn{0,x}}+B}\right.\\
& \left. \hspace{48pt} +\beta P^0\prn{x,x} C\prn{x}+\beta P^0\prn{x,x+1}C\prn{x+1}\right. \\
& \left. \hspace{102pt}+\beta P^0\prn{x,0}E\beta^{\tau\prn{0,x}}C\prn{x}-C\prn{x}\rule{0pt}{18pt} \right]\\
&\times \bra{1- \prn{\beta \prn{1-P^0\prn{x,0}} + \beta P^0\prn{x,0} E\beta^{\tau\prn{0,x}}}}^{-1}, \quad x \in \N/\cur{0}
\end{align*}
with
 \begin{align*}
 H\prn{0}&=\frac{ K\prn{0} +\beta P^0\prn{0,0} C\prn{0}+\beta P^0\prn{0,1}C\prn{1}-C\prn{0}}{1- \beta}.
\end{align*}

The following result is based in a self-consistency result for Gittins indices due to Nash \cite{Nash}:

\begin{lemma}\label{Nash}
 For the operation/intervention problem, if $H:\N \rightarrow \R$ is increasing, then $H\prn{x}\equiv G\prn{x},x \in \N$.
\end{lemma}

\begin{corollary}
 If $H\prn{x}$ is increasing, then $H\prn{x}=G\prn{x}$, $x \in \N$, and the $W$-index is given by
\begin{multline}\label{eq:WhitExp}
W\prn{x}=H\prn{x}\prn{1- \sum_{y=0}^{x-1}P^1\prn{x,y}\frac{ \prod_{h=y}^{x-1}\delta\prn{h}\prn{1-\sum_{i=0}^{y-1}\gamma\prn{i}\prod_{z=0}^{i-1}\delta\prn{z}}}{ 1-\sum_{i=0}^{x-1}\gamma\prn{i}\prod_{z=0}^{i-1}\delta\prn{z}}} \\
-\prn{\sum_{y=0}^{x-1}P^1\prn{x,y}\frac{ \prod_{h=y}^{x-1}\delta\prn{h}\prn{1-\sum_{i=0}^{y-1}\gamma\prn{i}\prod_{z=0}^{i-1}\delta\prn{z}}}{1-\sum_{i=0}^{x-1}\gamma\prn{i}\prod_{z=0}^{i-1}\delta\prn{z}}}C\prn{x} \\
-E_{y<x|x}\bra{K\prn{y,\tau\prn{y,x}}}
\end{multline}
\begin{proof}
The first statement is consequence of Lemma \ref{Nash}. The index $W\prn{x}$ follows directly from Theorem \ref{Theo:WhitMon}.
\end{proof}
\end{corollary}

\vspace{6pt}

\subsubsection*{Pure Deterioration Problem}

In certain cases, we may be interested in finding a maintenance scheduling regime that minimises the expected costs of operating the equipment without considering potential failures. In this \textit{pure deterioration} model, the machine is subject only to gradual deterioration. Consequently, Assumption \ref{ass:Probs} becomes:

\noindent\textbf{Assumption 4'.} Evolution under passivity is right skip free, namely
\[P^0\prn{x,x}+P^0\prn{x,x+1}=1,\quad x \in \N \]
All other characteristics of the model remain the same.

\cm{Notice that, in this case, it may be convenient to include a final absorbing state $\Delta$ that guarantees the machine's breakdown after certain level of wear (i.e. $P(\Delta,0)=1$). However, if the deterioration rate is slow enough and/or the number of states sufficiently large, the structure of the index policy guarantees that $\Delta$ will not be visited almost surely.}

Following a similar reasoning to the one used for the problem with failures, it is straightforward to derive a set of expressions equivalent to equations \eqref{beta_star} to \eqref{eq:Exp_K}. With these, we can construct an appropriate expression for quantity $\mathfrak{B}\prn{x,W}$ in \eqref{def:B} and, together with the definition of the index $G\prn{x}$ in \eqref{eq:Git}, we can define the $H\prn{x}$ function for the pure deterioration problem as
\begin{align}\label{eq:Nash_5}
 H\prn{x}=\frac{K\prn{x}+\beta P\prn{x,x}C\prn{x}+\beta P\prn{x,x+1}C\prn{x+1}-C\prn{x}}{1-\beta}
, \quad x \in \N \,.
\end{align}
The next result follows directly:
\begin{corollary}
 If $H\prn{x}$ is increasing, then $H\prn{x}=G\prn{x}$, $x \in \N$, and the $W$-index is given by 
\begin{multline}\label{eq:WhitInd}
W\prn{x}=\frac{K\prn{x}+\beta P\prn{x,x+1}\prn{C\prn{x+1}-C\prn{x}}}{1-\beta} \prn{1-\sum_{y=0}^{x-1}P^1\prn{x,y}\prod_{z=y}^{x-1}\delta \prn{z}} \\  
- \sum_{y=0}^{x-1}P^{1}\prn{x,y}\sum_{h=y}^{x-1}\kappa\prn{h}\prod_{z=y}^{h-1}\delta\prn{z} - C\prn{x}    , \quad x \in \N
\end{multline}
\begin{proof}
 The first statement is an immediate consequence of Lemma \ref{Nash}. The index $W\prn{x}$ in \eqref{eq:WhitInd}, follows from \eqref{eq:Whit}, the relevant expression for $\mathfrak{B}\prn{x,W}$ and \eqref{eq:Nash_5}.
\end{proof}
\end{corollary}

\noindent \textbf{Comment}

\dl{The reader may notice that quantity $\mathfrak{B}(x,W)$ in \eqref{def:B} is an conditional expectation on the state where the intervention is conducted. Alternatively, under perfect interventions equation (3) will become
 \begin{align*}
B\prn{x,W}+W=\frac{W+K\prn{\mb{0},\tau\prn{\mb{0},x}} +E\bra{\beta^{\tau\prn{\mb{0},x}}}C\prn{x} }{1-E\bra{\beta^{\tau\prn{\mb{0},x}}|\mb{0}}} \,.
\end{align*}
where $\mb{0}$ is the designated \textit{as-good-as-new} state. It can also be seen that expression $E_{y<x} K\prn{y,\tau(y,x)}$ in (21) becomes  $K\prn{x,\tau\prn{\mb{0},x}}$ under perfect interventions.  Consequently, the recursive computation of equations (20) is no longer necessary. This substantially simplifies the subsequent computations and the obtention of a closed form expression for the W-indices.}

\dl{Additionally, if $x$ is chosen as the arrival state after conducting the passive action (operation) during $\tilde{\tau}$ periods when starting from $\mb{0}$ (i.e. $\tau\prn{\mb{0},x}=\tilde \tau $), then $B\prn{\mb{0},W}$ becomes}
 \dl{\begin{align*}
B\prn{\mb{0},W}+W=\frac{W+K\prn{\mb{0},\tilde \tau} +E\bra{\beta^{\tilde{\tau}}}C\prn{X\prn{\tilde \tau}} }{1-E\bra{\beta^{\tilde{\tau}}|\mb{0}}} \,.
\end{align*} which is equivalent to expression (10) in \cite{GlaRuKi}. From here it is easy to see that the indices proposed in our manuscript are equivalent to the ones obtained by \cite{GlaRuKi} for the particular case of perfect intervention.}

\section{Numerical Assessment}\label{sec:Numerical}

In this section, we conduct a collection of numerical experiments in order to assess the performance of the proposed index policy. We start with a small illustrative example of the computation of the $W$-indices by means of the interaction between the proposed $\mathfrak{B}\prn{x,W}$ and $G\prn{x}$ functions. After this, we present the results of an extensive numerical comparison of the performance of the index policy with respect to an optimal stationary policy. Finally, given that the optimal policy can only be computed for relatively small examples, we conduct a simulation analysis of the relative efficiency of the index policy in larger systems with respect to alternative scheduling strategies that emerge in practice, namely, na\"ive and threshold policies.

\subsection{Computation of the W-Indices}

One of the main contributions of this article is the procedure for obtaining the $W$-index by means of the interaction of functions $\mathfrak{B}$ and $G$. As discussed in Section \ref{sec:Indices}, for a given state $x$ of the machine, the $W\prn{x}$ index is computed as the $W$-solution to the equation $G\prn{x}=\mathfrak{B}\prn{x,W}$. This procedure is illustrated with the following example.

Consider a system of four machines that are kept by two repairmen. The machines wear over a 24 states space with deterioration rates given by the vector $\prn{0.0208,\allowbreak  0.0245, \allowbreak  0.0180, \allowbreak 0.0149}$; with failure probability $P^0\prn{x,0}=0.0061e^{x/6},\allowbreak  \, x=1,\ldots, 24$. The transition probabilities for maintenance interventions are
\begin{align}\label{eq:ProbAct}
P^1\prn{x,y}=\frac{\pi\prn{x,y}}{\st \sum_{y<x}\pi\prn{x,y}}
\end{align}
 where $\pi\prn{x,y}=e^{-1.9436y}$. The intervention and operation costs are given by the functions $C_i\prn{x}=a_i+b_ix$ and $K\prn{x}=e_i+f_ix+g_ix^2$, respectively, and are expressed in monetary units. The corresponding parameters are shown in Table \ref{tab:param}. The failure cost is 4684.7 monetary units. All reported values of the parameters are random extractions from suitable uniform distributions (please see Section \ref{sec:subop} for details)\footnote{These values have been calibrated based on one of the author's experience on maintenance management and condition monitoring of wind turbines.}. However, it should be noticed that these values are provided only for illustrative purposes and any practical application of the results hereby presented requires a thorough analysis of the deteriorating process of the equipment under study.
 
\begin{figure}
 \begin{center}
  \begin{tabular}{cc}
   \fbox{\scalebox{0.3}{\includegraphics{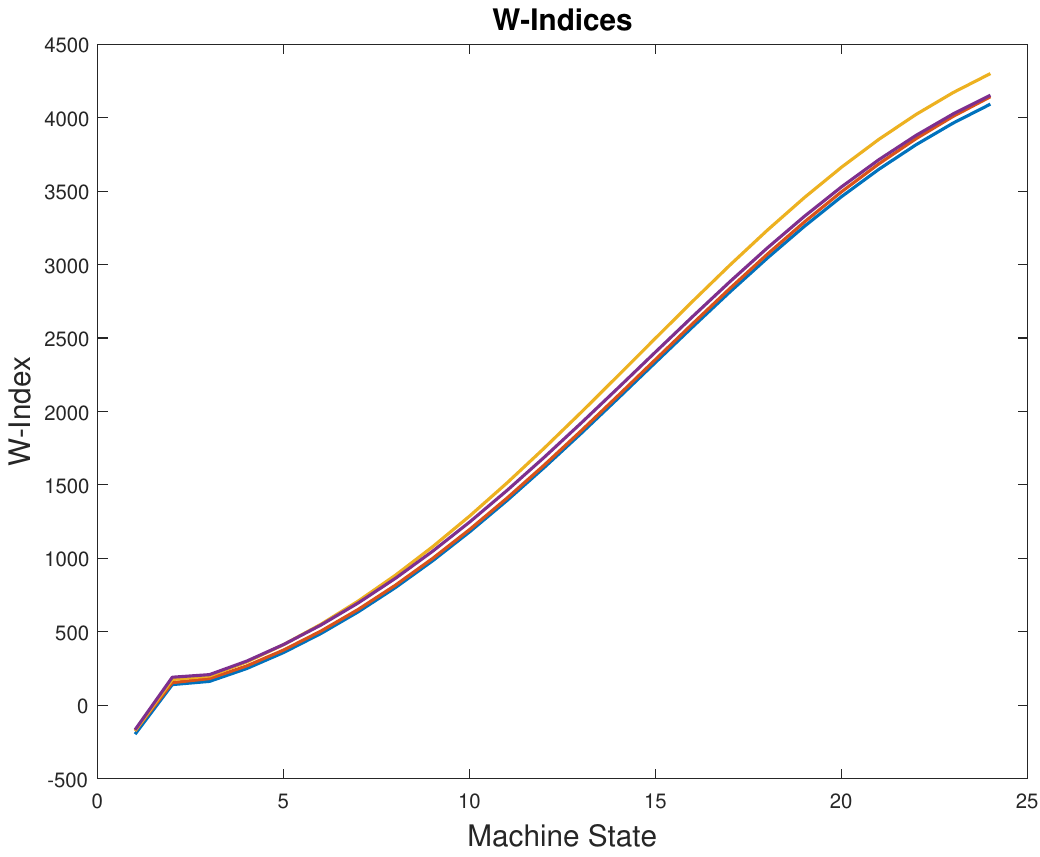}}} &
   \fbox{\scalebox{0.3}{\includegraphics{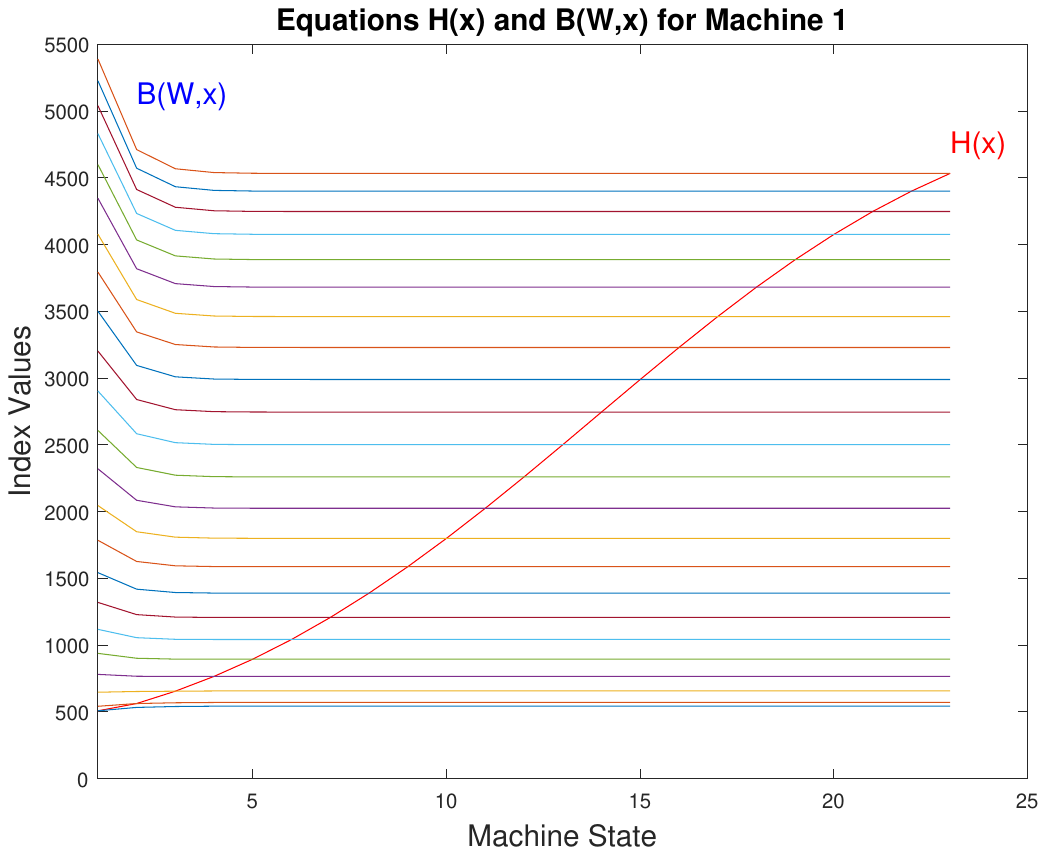}}}
  \end{tabular}
  \end{center}
   \caption{Determination of the W-indices}\label{fig:IndeDet}
\end{figure}

 The corresponding $W$-indices, computed using equation \eqref{eq:WhitExp}, are illustrated in the left hand side graph of Figure \ref{fig:IndeDet}. Each index is given by the intersection of $H\prn{x}$ with one of the level curves of $\mathfrak{B}$. This interaction is illustrated, for the case of Machine 1, in the right hand side of Figure \ref{fig:IndeDet}. Please notice that, given the fact that $\mathfrak{B}$ is continuous in $W$, the map of $\mathfrak{B}$ is dense, with one curve corresponding to each possible value of $W$. In Figure \ref{tab:param}, we have only highlighted those curves that correspond to the intersection between $\mathfrak{B}$ and $H$ for a given state $x$. For the sake of completeness, the corresponding $W$ indices are provided in Table \ref{tab:W-Ind}.

\begin{table}[hb]
 \begin{center}
\footnotesize
  \begin{tabular}{crrrr}\hline
 \st   Parameter & Mach. 1 & Mach. 2 & Mach. 3 & Mach. 4 \\ \hline
 \st  $a_i$ &174.5432& 197.2394  &  174.4626 & 166.8860\\
   $b_i$ & 11.8462&  14.5003    & 10.5560   &13.9013\\
   $e_i$ &  27.7880& 26.1096   & 24.2345 &  20.9082 \\
   $f_i$ &  1.3073& 1.5329    &  1.5620  &  1.8802 \\
   $g_i$ &  0.4915& 0.5054    &  0.5751  &  0.5036 \\
   \hline
  \end{tabular}
\caption{Cost Parameters for the Illustrative Example (monetary units)}\label{tab:param}
\end{center}
\end{table}

\begin{table}[ht]
\begin{center}
\footnotesize
 \begin{tabular}{ccccc} \hline
\st State  &	Machine 1& Machine 2 & Machine 3 & Machine 4 \\ \hline
\st 0	&	-197.24	&	-174.54	&	-174.46	&	-166.89	\\
1	&	140.28	&	153.50	&	174.89	&	190.53	\\
2	&	162.21	&	181.21	&	199.08	&	208.18	\\
3	&	249.40	&	267.25	&	294.24	&	300.03	\\
4	&	359.42	&	376.42	&	413.04	&	413.98	\\
5	&	488.43	&	505.04	&	551.54	&	546.18	\\
6	&	635.71	&	652.10	&	709.06	&	696.22	\\
7	&	800.78	&	817.02	&	885.08	&	863.75	\\
8	&	982.94	&	999.09	&	1078.75	&	1048.06	\\
9	&	1181.03	&	1197.20	&	1288.72	&	1247.98	\\
10	&	1393.41	&	1409.77	&	1513.15	&	1461.87	\\
11	&	1617.95	&	1634.72	&	1749.66	&	1687.55	\\
12	&	1851.94	&	1869.41	&	1995.34	&	1922.33	\\
13	&	2092.26	&	2110.76	&	2246.80	&	2163.07	\\
14	&	2335.37	&	2355.28	&	2500.31	&	2406.23	\\
15	&	2577.51	&	2599.25	&	2751.91	&	2648.09	\\
16	&	2814.87	&	2838.87	&	2997.64	&	2884.85	\\
17	&	3043.76	&	3070.45	&	3233.69	&	3112.87	\\
18	&	3260.83	&	3290.63	&	3456.67	&	3328.83	\\
19	&	3463.23	&	3496.54	&	3663.72	&	3529.94	\\
20	&	3648.75	&	3685.92	&	3852.67	&	3714.05	\\
21	&	3815.88	&	3857.20	&	4022.11	&	3879.67	\\
22	&	3963.81	&	4009.53	&	4171.36	&	4026.08	\\
23	&	4092.41	&	4142.72	&	4300.43	&	4153.17	\\
24	&	4132.57	&	4187.48	&	4352.13	&	4188.39	\\
\hline
 \end{tabular}
 \caption{Values of the $W$-Indices for the illustrative example}\label{tab:W-Ind}
\end{center}
\end{table}

\subsection{Suboptimality Assessment of the Index Policy}\label{sec:subop}

\cm{In the following lines we assess the effectivity of the index policy by comparing its performance with the optimal stationary policy described in the discussion around equation \eqref{eq:MDP}.} As it was mentioned before, the problem of scheduling maintenance interventions to a collection of deteriorating machines -as formulated in this article- suffers from the so-called curse of dimensionality. This dimensionality problem hinders the application of standard dynamic programming techniques to the solution of the associated MDP problem. For this reason, the numerical assessment in this section is limited to small sized instances of the problem.

We explore a scenario where two repairmen are in charge of maintaining five machines. \cm{We further introduce two additional \textit{virtual} machines to account for situations when the system will find efficient to keep one or both technicians idle. For example, if the machines are in early stages of wear and intervention is not necessary.}

The values of the W-indices for the corresponding machine-state pairs are computed, by means of equation \eqref{eq:WhitExp}, with parameters calibrated as follows: The deterioration rates are obtained from a $U\prn{0.01,0.025}$ distribution. The failure probabilities are given by the expression $P\prn{x,0}=qe^{x/4}$, where the $q$ values follow a $U\prn{0.005,0.015}$ distribution. The transition probabilities under intervention are given by expression \eqref{eq:ProbAct}, where $\pi\prn{x,y}=e^{-\nu y}$ and $\nu$ is obtained from a $U\prn{0,2}$ distribution. A linear configuration has been considered for the maintenance costs, with five possible distributions for the constant term: $U\prn{50,80},U\prn{100,140},U\prn{250,300},U\prn{500,600}$ and $U\prn{1000,1200}$; and the slopes following a $U\prn{5,15}$ distribution. Regarding the operation costs, we adopt the common practice of assuming increasing costs (see, for example \cite{ValFel1989,GriVanSpi2006,liuWuXie2017}  and consider two different configurations: linear and quadratic.  For the linear configuration we have four possible scenarios given by the combination of two alternative distributions for the intercept, namely $U\prn{20,30}$ and $U\prn{40,60}$; and two distributions for the slope, $U\prn{1,3}$ and $U\prn{8,12}$, respectively. For the quadratic configuration, a quadratic term is added to the four linear cases, with the corresponding coefficient obtained from a $U\prn{0.4,0.6}$ distribution. The failure costs take random values between 7.5 and 12.5 times the average intervention costs on each possible case. The value of the index for the two virtual machines is, trivially, zero.

The experiments were conducted as follows: 1000 problems were solved for each pair of configurations of the intervention and operation costs (linear-linear and linear-quadratic). There are 20 scenarios in each configuration and 50 experiments were conducted for each scenario. Those 50 problems differ in the values of the parameters for the operation costs and the transition probabilities, which are computed as described above. In total, 2000 experiments were conducted. As indicated by Lemma \ref{Nash}, a particular problem is indexable if and only if function $H\prn{x}$ in increasing in $x$; accordingly, this condition was tested for every set of randomly generated parameters before running the experiment. \cm{For each experiment, the optimal expected discounted cost (with a tolerance parameter $\epsilon=0.01\%$) and the expected discounted cost of the index policy, were obtained by means of a standard value iteration algorithm (see for example \cite{Ross,Puterman}). Averages over 1000 repetitions were taken. With these values, we obtained the percentage-cost suboptimality of the index policy.}

The results of the numerical assessment are summarised in Table \ref{tab:Case_1}. Each row shows the order statistics for the corresponding operation/intervention costs configuration. The label ``Case'' in the Maintenance Costs column refers to each of the five proposed distributions of the constant parameter in the linear maintenance cost function.

\begin{table}[hbt]
 \begin{center}
\footnotesize
  \begin{tabular}{c|c|ccccc}\hline
\st Operation       & Maint.  &         & Lower    &        & Upper    & \\
\st Costs  & Costs & Minimum & Quartile & Median & Quartile & Maximum \\\hline
\st \multirow{5}{*}{Linear}	&	Case 1	&	0.000	&	0.001	&	0.012	&	0.061	&	0.169	\\
	&	Case 2	&	0.001	&	0.015	&	0.063	&	0.192	&	0.367	\\
	&	Case 3	&	0.025	&	0.141	&	0.226	&	0.347	&	0.869	\\
	&	Case 4	&	0.131	&	0.333	&	0.468	&	0.683	&	1.014	\\
	&	Case 5	&	0.626	&	0.848	&	1.127	&	1.370	&	2.128	\\ \hline

\st \multirow{5}{*}{Quadratic}	&	Case 1	&	0.000	&	0.000	&	0.004	&	0.015	&	0.087	\\
	&	Case 2	&	0.000	&	0.009	&	0.030	&	0.063	&	0.216	\\
	&	Case 3	&	0.019	&	0.095	&	0.131	&	0.328	&	0.734	\\
	&	Case 4	&	0.156	&	0.307	&	0.398	&	0.540	&	0.946	\\
	&	Case 5	&	0.298	&	0.895	&	1.317	&	2.019	&	3.429	\\ \hline

\end{tabular}
     \caption{Performance of the Index policy with failures (\% suboptimality)}\label{tab:Case_1}
 \end{center}
\end{table}

It can be seen that the suboptimality of the index policy was never above $3.5\%$ of the optimal policy. The apparent increase in suboptimality for large costs scenarios may be explained by the fact that the index policy tends to intervene less than the optimal policy when the intervention costs increase (a larger number of states will present negative indices), therefore incurring in larger failure costs.

For the sake of completeness, the experiments were also conducted for the pure deterioration case. We recall that in this case, machines are only subject to gradual deterioration and failures never occur. Therefore, the parameters associated to the failures, namely, $B,P^0\prn{x,0}$ and $\gamma\prn{x}$, are all set equal to zero. Additionally, the $W$-indices are computed according to equation \eqref{eq:WhitInd}. Table \ref{tab:Case_3} summarises our findings. 

\begin{table}[hbt]
 \begin{center}
\footnotesize
  \begin{tabular}{c|c|ccccc}\hline
\st Operation& Maint. 		&         & Lower    &        & Upper    & \\
\st Costs  & Costs   	& Minimum & Quartile & Median & Quartile & Maximum \\\hline
\st \multirow{5}{*}{Linear}	&	Case 1	&	0.000	&	0.263	&	0.589	&	1.245	&	6.324	\\
	&	Case 2	&	0.000	&	0.000	&	0.077	&	0.619	&	1.505	\\
	&	Case 3	&	0.000	&	0.000	&	0.000	&	0.314	&	1.155	\\
	&	Case 4	&	0.000	&	0.000	&	0.000	&	0.000	&	0.438	\\
	&	Case 5	&	0.000	&	0.000	&	0.000	&	0.000	&	0.000	\\ \hline
\st \multirow{5}{*}{Quadratic}	&	Case 1	&	0.000	&	0.401	&	0.692	&	1.553	&	5.651	\\
	&	Case 2	&	0.000	&	0.281	&	0.526	&	0.820	&	1.432	\\
	&	Case 3	&	0.000	&	0.000	&	0.000	&	0.183	&	0.685	\\
	&	Case 4	&	0.000	&	0.000	&	0.000	&	0.116	&	0.624	\\
	&	Case 5	&	0.000	&	0.000	&	0.000	&	0.000	&	0.000	\\ \hline

     \end{tabular}
     \caption{Performance of the Index policy for the pure-deterioration case (\% suboptimality)}\label{tab:Case_3}
 \end{center}
\end{table}

In this case, the maximum suboptimality is above $6\%$ with respect to the optimal policy. However, the results suggest a much stronger performance of the index policy when the system is not subject to failures than with them. Indeed, for large values of the intervention costs, the index policy tends to behave optimally. This may be the case because the index policy tends to intervene machines at an earlier state of wear than the optimal policy (notice that the index policy will always intervene in a machine with a positive index value), increasing the total intervention costs. As the maintenance cost increases, the number of states with negative index value becomes larger, and the number of interventions prescribed by index policy in early states of wear, decreases. This reduces the total intervention costs bringing the value of the index policy closer to the one optimal one.

\subsection{Index Policy Performance in Large Systems}

For large sized instances, dynamic programming techniques are no longer available for finding the optimal solution to the problem due to prohibitively expensive computational costs. In those cases, the performance of the index policy must be assessed by comparison with other alternative policies. In this section, we conduct a series of simulation experiments where the efficiency of the index policy proposed in this article is compared with two alternative policies: myopic and threshold.

Two sets of simulations were conducted, one with 2 repairmen and 25 machines, and the other one with 3 repairmen and 50 machines. The W-indices are computed according to equation \eqref{eq:WhitExp} and the parameters calibrated in the way described in Section \ref{sec:subop}. We considered four alternative distributions for the constant term in the linear intervention cost, namely $U\prn{250,295},U\prn{500,560}$, $U\prn{750,825}$ and $U\prn{1000,1090}$; with the slope given by a $U\prn{10,15}$ distribution. The operation costs have a linear configuration with the constant term following either a $U\prn{20,30}$ or a $\prn{40,60}$ distribution, and the slope distributed as either a $U\prn{1,3}$ or a $U\prn{8,12}$. The failure costs are computed as the mean intervention cost times a value taken randomly from one of the following intervals $\prn{5,15},\prn{15,25}$ and $\prn{50,60}$.  The deterioration rates are obtained from a $U\prn{0.01,0.025}$ distribution. The failure probabilities are given by the expression $P\prn{x,0}=qe^{x/6}$, where the $q$ values follow a $U\prn{0.0025,0.0050}$ distribution. The transition probabilities under intervention are given by expression \eqref{eq:ProbAct}, where $\pi\prn{x,y}=e^{-\nu y}$ and $\nu$ is obtained from a $U\prn{0,2}$ distribution. As before, the indexability of the problem is tested for each new combination of parameters.

Simulations were conducted for each of 48 possible combinations of the different intervention, operation and failure costs scenarios. A total of 250 simulations, differing in the operation costs and transition probabilities, were conducted for each of those combinations. In total 12000 simulations were ran. For each simulation we computed the average expected discounted value of the index policy, together with the corresponding values of a myopic policy, which intervenes the most deteriorated machines at each decision epoch\footnote{\cm{We refer the reader to the end of Section \ref{sec:Indices} for a discussion on the different ordering of the states resulting from the index and the na\"ive policies.}}; and a set of eight threshold policies, intervening whenever a machine arrives to certain pre-determined state, with ties broken randomly. Each simulation was taken over an assumed planning horizon of 520 decision periods. Table \ref{tab:Sim_1} shows the results of the simulations for the case with 25 machines and 2 repairmen; Table \ref{tab:Sim_2} illustrates the case with 50 machines and 3 repairmen. 

\cm{These tables include the average expected discounted cost of each policy, and the average of the expected discounted value of the best threshold policy in each scenario. It also includes the average number of interventions and failures corresponding to each policy. As before, the ``Case'' labels in the tables correspond to each of the four possible distributions of the constant term for the intervention costs. The three possible cases of failure costs are labelled \textit{Low}, \textit{Medium}, and \textit{High}.}

\begin{sidewaystable}
\footnotesize
\centering
\begin{tabular}{c|c|rrr|rrr|rrrr|rr}\hline
\st  Failure & Maint.& \multicolumn{3}{c|}{Index Policy}  & \multicolumn{3}{c|}{Na\"ive Policy} & \multicolumn{4}{c}{Threshold Policy} & \multicolumn{2}{c}{Performance Ratios} \\ \cline{3-14}
\st Costs	   & Costs& Cost  & Intervs.  & Failures  & Cost  & Intervs.  & Failures  & Cost & Intervs. & Failures & Best & Index/Na\"ive & Index/Threshold\\ \hline
\st \multirow{4}{*}{Low}	&	Case 1	&	14423.1	&	618.9	&	64.7	&	14825.0	&	936.7	&	43.6	&	14742.5	&	578.1	&	71.7	&	14571.3	& 2.7 \% & 2.2 \% \\
	&	Case 2	&	16372.1	&	434.0	&	82.7	&	18578.4	&	935.6	&	43.6	&	17282.9	&	576.9	&	72.2	&	16381.0	& 11.9 \% & 5.3 \%\\
	&	Case 3	&	17410.6	&	357.0	&	89.4	&	21660.8	&	935.9	&	43.4	&	19188.6	&	577.3	&	71.7	&	17367.8	& 19.6 \% & 9.3 \% \\
	&	Case 4	&	21538.1	&	367.7	&	89.1	&	26894.1	&	936.0	&	43.4	&	23734.5	&	577.2	&	71.9	&	21425.4	& 19.9 \% & 9.3 \% \\ \hline
\st \multirow{4}{*}{Medium}	&	Case 1	&	16122.3	&	837.2	&	47.7	&	16369.7	&	935.7	&	43.2	&	16705.7	&	576.7	&	71.8	&	15796.5	& 1.5 \% & 3.5 \% \\
	&	Case 2	&	21129.4	&	674.7	&	59.0	&	21740.4	&	936.0	&	42.7	&	21457.5	&	577.7	&	71.0	&	21197.4	& 2.8 \% & 1.5 \% \\
	&	Case 3	&	26275.0	&	651.1	&	61.3	&	27477.4	&	934.4	&	43.3	&	26772.6	&	576.4	&	72.0	&	26259.8	& 4.4 \% & 1.9 \% \\
	&	Case 4	&	29329.8	&	563.0	&	69.5	&	32146.6	&	937.0	&	44.0	&	30391.7	&	577.6	&	72.9	&	29296.9	& 8.8 \% & 3.5 \% \\ \hline
\st \multirow{4}{*}{High}	&	Case 1	&	19034.7	&	1039.3	&	33.7	&	22885.5	&	936.9	&	43.2	&	25056.6	&	577.4	&	71.6	&	19464.1	& 16.8 \% & 24.0 \% \\
	&	Case 2	&	26958.1	&	1039.3	&	33.6	&	32697.5	&	933.7	&	43.4	&	35359.2	&	575.8	&	72.0	&	26945.6	& 17.6 \% & 23.8 \% \\
	&	Case 3	&	35343.2	&	1039.2	&	33.5	&	43497.3	&	934.7	&	43.5	&	47188.5	&	576.7	&	71.8	&	36252.8	& 18.7 \% & 25.1 \% \\
	&	Case 4	&	42266.0	&	1036.6	&	33.3	&	50900.5	&	934.7	&	43.3	&	55077.0	&	576.2	&	71.7	&	43381.0	& 17.0 \% & 23.3 \% \\ \hline

\end{tabular}
\caption{Simulation results for 25 machines and 2 repairmen}\label{tab:Sim_1} 

\vspace{18pt}

\footnotesize
\centering
\begin{tabular}{c|c|rrr|rrr|rrrr|rr}\hline
\st  Failure & Maint.& \multicolumn{3}{c|}{Index Policy}  & \multicolumn{3}{c|}{Na\"ive Policy} & \multicolumn{4}{c}{Threshold Policy} & \multicolumn{2}{c}{Performance Ratios} \\ \cline{3-14}
\st Costs	   & Costs& Cost  & Intervs.  & Failures  & Cost  & Intervs.  & Failures  & Cost & Intervs. & Failures & Best & Index/Na\"ive & Index/Threshold\\ \hline
\st \multirow{4}{*}{Low}	&	Case 1	&	28190.9	&	1167.3	&	133.9	&	28982.9	&	1548.0	&	106.7	&	28692.8	&	1044.3	&	150.6	&	28370.1	& 2.7 \% & 1.7 \% \\
	&	Case 2	&	31897.2	&	835.7	&	169.2	&	35981.0	&	1548.2	&	107.2	&	33502.4	&	1044.9	&	150.8	&	31946.8	& 11.3 \% & 4.8 \% \\
	&	Case 3	&	36532.3	&	734.2	&	179.1	&	43716.7	&	1548.0	&	107.3	&	39392.1	&	1045.1	&	150.9	&	36546.7	& 16.4 \% & 7.3 \% \\
	&	Case 4	&	43837.4	&	781.4	&	169.3	&	53141.0	&	1547.9	&	107.2	&	47393.8	&	1044.3	&	151.1	&	43637.9	& 17.5 \% & 7.5 \% \\ \hline
\st \multirow{4}{*}{Medium}	&	Case 1	&	31757.6	&	1405.6	&	116.0	&	32142.7	&	1548.2	&	107.2	&	32692.9	&	1044.3	&	150.8	&	31862.3	& 1.2 \% & 2.9 \% \\
	&	Case 2	&	41331.5	&	1272.9	&	123.9	&	42630.3	&	1548.1	&	106.3	&	41835.5	&	1044.4	&	149.9	&	41291.8	& 3.0 \% & 1.2 \% \\
	&	Case 3	&	52774.7	&	1239.3	&	127.4	&	54747.9	&	1548.1	&	107.2	&	53396.9	&	1044.7	&	151.1	&	52590.7	& 3.6 \% & 1.2 \% \\
	&	Case 4	&	56066.9	&	1069.8	&	140.3	&	61618.7	&	1548.2	&	106.3	&	58022.1	&	1045.1	&	149.9	&	55973.0	& 9.0 \% & 3.4 \% \\ \hline
\st \multirow{4}{*}{High}	&	Case 1	&	41917.8	&	1559.9	&	105.0	&	46096.0	&	1548.2	&	106.3	&	50782.0	&	1044.6	&	149.8	&	42301.9	& 9.1 \% & 17.5 \% \\
	&	Case 2	&	60614.0	&	1559.9	&	104.7	&	66547.3	&	1548.2	&	106.4	&	72823.6	&	1045.4	&	150.1	&	60956.4	& 8.9 \% & 16.8 \% \\
	&	Case 3	&	76506.9	&	1559.8	&	104.9	&	84357.1	&	1547.9	&	107.0	&	91582.3	&	1044.5	&	150.4	&	76928.7	& 9.3 \% & 16.5 \% \\
	&	Case 4	&	97177.9	&	1559.8	&	105.0	&	109199.2	&	1548.1	&	107.0	&	118016.4	&	1044.8	&	150.6	&	99279.3	& 11.0 \% & 17.7 \% \\ \hline
\end{tabular}
\caption{Simulation results for 50 machines and 3 repairmen}\label{tab:Sim_2}
\end{sidewaystable}

It can be seen that the index policy outperforms both the myopic policy and the average threshold policies in all cases. \cm{The best threshold policy outperforms the index one in a number of cases, but the fact that the index policy is better in all the others gives an idea of the strength of the proposed heuristic. This becomes evident when we consider that choosing the best threshold policy for a particular scenario happens only by chance or, at best, after a very large learning process, over which the decision maker will be incurring the -larger- average cost of the threshold policy. Other results, which are consistent with what would have been intuitively expected, are: (a) as the maintenance cost increases, the number of interventions prescribed by the index policy decreases; (b) as the failure costs increase, so does the number of interventions prescribed by the index heuristic; (c) in average the index policy tends to intervene less than the myopic policy and more than the threshold policy; and (d) given that the myopic policy does not consider costs, it tends to intervene more than necessary, therefore the smaller number of failures comparing with the other two policies; however, this comes at a much larger overall cost.}

\subsection{\cm{An alternative index policy}}

\cm{In this section we make a numerical comparison of the performance of our proposed index policy against the one proposed in \cite{GlaRuKi}. This alternative policy, to which we refer to as \textit{myopic}, ignores the fact that errors or imperfections may arise when conducting preventive maintenance, and computes the indices assuming that the interventions are perfect, i.e. they return the machine to its \textit{as-good-as-new} state. Notwithstanding both policies are index-based, for the sake of consistency with the rest of the manuscript, we retain the name \textit{Index Policy} for the one proposed in this work.}

\cm{Table \ref{tab:myo} shows the simulation results for a problem with 50 machines and 3 technicians. Simulations consisted of running 250 instances of each of 12 possible configurations following the scheme used in the previous section. The results show that, apart from a couple of instances with low maintenance and failures costs, the \textit{Index Policy} outperforms the myopic one. The results also suggest that the myopic policy tends to intervene more than the index policy, but the interventions seem to be less efficient. This becomes more evident when the failure cost is high, and the number of interventions required to minimise the operation/intervention costs is large. In those cases, the number of failures experienced when using the myopic policy is larger.}

\begin{table}[h!]
\centering
\footnotesize
\begin{tabular}{|c|c|rrr|rrr|}
 \hline																
\st Failure	&	Maint.	&	\multicolumn{3}{|c|}{Index Policy}			&			\multicolumn{3}{|c|}{Myopic Policy}			\\	\cline{3-8}
\st Costs	&	Costs	&	Cost	&	Intervs.	&	Failures	&	Cost	&	Intervs.	&	Failures	\\	\hline
\st Low	&	Case 1	&	21918.0	&	638.1	&	141.1	&	21912.8	&	847.3	&	121.5	\\	
	&	Case 2	&	19584.7	&	230.0	&	221.4	&	19587.6	&	267.5	&	209.2	\\	
	&	Case 3	&	31326.2	&	447.1	&	169.0	&	31338.2	&	546.7	&	154.2	\\	
	&	Case 4	&	36769.8	&	424.4	&	170.4	&	36770.7	&	515.1	&	156.1	\\	\hline
\st Medium	&	Case 1	&	24126.4	&	820.6	&	125.1	&	24073.2	&	1119.9	&	103.7	\\	
	&	Case 2	&	35901.6	&	876.5	&	117.9	&	35919.8	&	1184.5	&	96.7	\\	
	&	Case 3	&	42970.7	&	768.1	&	129.2	&	43045.6	&	1023.7	&	109.3	\\	
	&	Case 4	&	54193.7	&	798.5	&	126.5	&	54495.6	&	1071.3	&	106.1	\\	\hline
\st High	&	Case 1	&	36596.7	&	1559.9	&	73.1	&	36717.7	&	1559.9	&	74.5	\\	
	&	Case 2	&	53147.2	&	1559.9	&	72.0	&	53315.7	&	1559.9	&	73.8	\\	
	&	Case 3	&	69988.1	&	1559.9	&	72.7	&	70174.0	&	1559.9	&	73.7	\\	
	&	Case 4	&	85219.3	&	1559.9	&	71.2	&	85313.1	&	1559.9	&	73.1	\\	\hline

\end{tabular}
\caption{Simulation results for the Index and Myopic policies.}\label{tab:myo}
\end{table}

\section{Conclusions}\label{sec:Conc}

In this article, we analyse the problem of scheduling maintenance interventions for a collection of deteriorating machines by a limited number of repairmen. We acknowledge that, in real life, maintenance interventions are subject to errors (induced either by human factors or by unforeseeable external events), which may imply that the final state of the intervened machine differs from the pursued -or planned- one. In order to account for this fact, we propose an MDP formulation of the machine maintenance problem with imperfect interventions. Given the inherent complexity of the problem, whose size grows exponentially for real-life sized instances, we propose a restless-bandit formulation that exploits the mathematical characteristics of the individual machines in order to develop an index-based scheduling policy. Based on the dynamics of a deteriorating machine, we propose an index structure and, after establishing an indexability property, we develop closed form indices for the individual machines, which build on the so-called Whittle indices for restless bandit problems.

The performance of the proposed index policy is assessed in two different sets of experiments. For small instances, where dynamic programming techniques are available for finding the optimal solution to the problem, a set of experiments were conducted in order to evaluate the performance of the index policy. The experimental results suggest a very strong performance of the index heuristic, with maximum suboptimality below $3.5\%$ for very large intervention costs and below $1\%$ in most cases. In a number of cases, the index policy was found to perform optimally. For larger instances, the index policy was compared with two alternative policies: a so-called na\"ive one, which always intervenes in the machines with more advanced state of wear (irrespectively of the involved costs); and a threshold policy, which prescribes intervention whenever the machine reaches certain predetermined state of wear. The index policy outperforms both policies in a series of simulations conducted in two scenarios with 25 and 50 machines, respectively. This result suggests a very strong performance of the proposed heuristic.

\cm{Additionally, in order to highlight the importance of taking into account the existence of imperfections in the preventive maintenance tasks, we compare the results of our heuristic against an index policy that assumes perfect interventions. Numerical results show that the policy that accounts for imperfections outperforms the so-called myopic policy in most instances. This becomes more important as the maintenance and failure costs increase.}

Our work has extended the available literature in machine maintenance by proposing an intuitive and efficient mechanism for allocating effort among deteriorating machines. This approach can be combined with advanced condition monitoring and predictive maintenance systems in order to obtain an accurate characterisation of the system's deterioration and wear. Moreover, the technique proposed in this article opens new research and implementation opportunities, as it can be used in combination with other techniques available for optimal maintenance in single machines, in order to guarantee a system's availability, reliability, and productivity at a minimal cost.

\bibliographystyle{apalike}
\bibliography{BanditBib}

\begin{thebibliography}{}

\bibitem[Abad and Iyengar, 2016]{AbIy2016}
Abad, C. and Iyengar, G. (2016).
\newblock A near-optimal maintenance policy for automated {DR} devices.
\newblock {\em IEEE Transactions on Smart Grid}, 7(3):1411--1419.

\bibitem[Alaswad and Xiang, 2017]{AlaXia2017}
Alaswad, S. and Xiang, Y. (2017).
\newblock A review on condition-based maintenance optimization models for
  stochastically deteriorating system.
\newblock {\em Reliability Engineering \& System Safety}, 157:54--63.

\bibitem[Ansell et~al., 2003]{KG02}
Ansell, P., Glazebrook, K., Ni{\~n}o-Mora, J., and O'Keeffe, M. (2003).
\newblock Whittle's index policy for a multi-class queueing system with convex
  holding costs.
\newblock {\em Mathematical Methods of Operations Research}, 57:21--39.

\bibitem[Ben-Daya et~al., 2009]{BenDufRao2009}
Ben-Daya, M., Ait-Kadi, D., Duffuaa, S., Knezevic, J., and Raouf, A. (2009).
\newblock {\em Handbook of maintenance management and engineering}, volume~7.
\newblock Springer.

\bibitem[Borrero and Akhavan-Tabatabaei, 2013]{BorAkh2013}
Borrero, J.~S. and Akhavan-Tabatabaei, R. (2013).
\newblock Time and inventory dependent optimal maintenance policies for single
  machine workstations: An {MDP} approach.
\newblock {\em European Journal of Operational Research}, 228(3):545--555.

\bibitem[Canto, 2008]{Canto2008}
Canto, S.~P. (2008).
\newblock Application of {B}ender{'}s decomposition to power plant preventive
  maintenance scheduling.
\newblock {\em European journal of operational research}, 184(2):759--777.

\bibitem[Cao et~al., 2018]{CaoJiaHu2018}
Cao, W., Jia, X., Hu, Q., Zhao, J., and Wu, Y. (2018).
\newblock A literature review on selective maintenance for multi-unit systems.
\newblock {\em Quality and Reliability Engineering International},
  34(5):824--845.

\bibitem[Detti et~al., 2019]{DetNicPac2019}
Detti, P., Nicosia, G., Pacifici, A., and de~Lara, G. Z.~M. (2019).
\newblock Robust single machine scheduling with a flexible maintenance
  activity.
\newblock {\em Computers \& Operations Research}.

\bibitem[Do~Van and B{\'e}renguer, 2012]{VanBer2012}
Do~Van, P. and B{\'e}renguer, C. (2012).
\newblock Condition-based maintenance with imperfect preventive repairs for a
  deteriorating production system.
\newblock {\em Quality and Reliability Engineering International},
  28(6):624--633.

\bibitem[Do~Van et~al., 2015]{DoVoiLev2015}
Do~Van, P., Voisin, A., Levrat, E., and Iung, B. (2015).
\newblock A proactive condition-based maintenance strategy with both perfect
  and imperfect maintenance actions.
\newblock {\em Reliability Engineering \& System Safety}, 133:22--32.

\bibitem[Froger et~al., 2017]{FroGenMen2017}
Froger, A., Gendreau, M., Mendoza, J.~E., Pinson, E., and Rousseau, L.-M.
  (2017).
\newblock A branch-and-check approach for a wind turbine maintenance scheduling
  problem.
\newblock {\em Computers \& Operations Research}, 88:117--136.

\bibitem[Gharbi and Kenn{\'e}, 2005]{GhaKen2005}
Gharbi, A. and Kenn{\'e}, J.-P. (2005).
\newblock Maintenance scheduling and production control of multiple-machine
  manufacturing systems.
\newblock {\em Computers \& industrial engineering}, 48(4):693--707.

\bibitem[Gilardoni et~al., 2016]{GilTolFre2016}
Gilardoni, G., {Guerra-de-Toledo}, M., Freitas, M., and Colosimo, E. (2016).
\newblock Dynamics of an optimal maintenance policy for imperfect repair
  models.
\newblock {\em European Journal of Operational Research}, 248(3):1104--1112.

\bibitem[Gittins, 1979]{Gittins}
Gittins, J. (1979).
\newblock Bandit processes and dynamic allocation indices.
\newblock {\em Journal of the Royal Statistical Society, Series B},
  41:148--177.

\bibitem[Gittins et~al., 2011]{GitGlaWe2011}
Gittins, J., Glazebrook, K., and Weber, R. (2011).
\newblock {\em Multi-armed bandit allocation indices}.
\newblock John Wiley \& Sons.

\bibitem[Gittins and Jones, 1979]{GitJon}
Gittins, J. and Jones, D. (1979).
\newblock A dynamic allocation index for the discounted multiarmed bandit
  problem.
\newblock {\em Biometrika}, 66:561--565.

\bibitem[Glazebrook et~al., 2014]{GlaHoKi2014}
Glazebrook, K., Hodge, D., Kirkbride, C., and Minty, R. (2014).
\newblock Stochastic scheduling: A short history of index policies and new
  approaches to index generation for dynamic resource allocation.
\newblock {\em Journal of Scheduling}, 17(5):407--425.

\bibitem[Glazebrook et~al., 2005]{GlaMitAns}
Glazebrook, K., Mitchell, H., and Ansell, P. (2005).
\newblock Index policies for the maintenance of a collection of machines by a
  set of repairmen.
\newblock {\em European Journal of Operational Research}, 165:267--284.

\bibitem[Glazebrook et~al., 2002]{KG01}
Glazebrook, K., Ni{\~n}o-Mora, J., and Ansell, P. (2002).
\newblock Index policies for a class of discounted restless bandits.
\newblock {\em Adv. Appl. Prob.}, 34:754--774.

\bibitem[Glazebrook et~al., 2006]{GlaRuKi}
Glazebrook, K., Ruiz-Hernandez, D., and Kirkbride, C. (2006).
\newblock Some indexable families of restless bandit problems.
\newblock {\em Advances in Applied Probability}, 38:643--672.

\bibitem[Grall et~al., 2002]{GraBerDie2002}
Grall, A., B{\'e}renguer, C., and Dieulle, L. (2002).
\newblock A condition-based maintenance policy for stochastically deteriorating
  systems.
\newblock {\em Reliability Engineering \& System Safety}, 76(2):167--180.

\bibitem[Grigoriev et~al., 2006]{GriVanSpi2006}
Grigoriev, A., Van De~Klundert, J., and Spieksma, F.~C. (2006).
\newblock Modeling and solving the periodic maintenance problem.
\newblock {\em European Journal of Operational Research}, 172(3):783--797.

\bibitem[Hsu et~al., 2013]{HsuJiGuo2013}
Hsu, C.-J., Ji, M., Guo, J.-Y., and Yang, D.-L. (2013).
\newblock Unrelated parallel-machine scheduling problems with aging effects and
  deteriorating maintenance activities.
\newblock {\em Information Sciences}, 253:163--169.

\bibitem[Huynh et~al., 2012]{HuyCasBar2012}
Huynh, K.~T., Castro, I.~T., Barros, A., and B{\'e}renguer, C. (2012).
\newblock Modeling age-based maintenance strategies with minimal repairs for
  systems subject to competing failure modes due to degradation and shocks.
\newblock {\em European journal of operational research}, 218(1):140--151.

\bibitem[Irawan et~al., 2017]{IraOueJon2017}
Irawan, C.~A., Ouelhadj, D., Jones, D., St{\aa}lhane, M., and Sperstad, I.~B.
  (2017).
\newblock Optimisation of maintenance routing and scheduling for offshore wind
  farms.
\newblock {\em European Journal of Operational Research}, 256(1):76--89.

\bibitem[Kenne et~al., 2003]{KenBouGha2003}
Kenne, J., Boukas, E., and Gharbi, A. (2003).
\newblock Control of production and corrective maintenance rates in a
  multiple-machine, multiple-product manufacturing system.
\newblock {\em Mathematical and Computer Modelling}, 38(3):351--365.

\bibitem[Khatab and Aghezzaf, 2016]{KhaAgh2016}
Khatab, A. and Aghezzaf, E.-H. (2016).
\newblock Selective maintenance optimization when quality of imperfect
  maintenance actions are stochastic.
\newblock {\em Reliability Engineering \& System Safety}, 150:182--189.

\bibitem[Kobbacy and Murthy, 2008]{KobMur2008}
Kobbacy, K. A.~H. and Murthy, D.~P. (2008).
\newblock {\em Complex system maintenance handbook}.
\newblock Springer Science \& Business Media.

\bibitem[Kurt and Kharoufeh, 2010]{KurKha2010}
Kurt, M. and Kharoufeh, J.~P. (2010).
\newblock Optimally maintaining a markovian deteriorating system with limited
  imperfect repairs.
\newblock {\em European Journal of Operational Research}, 205(2):368--380.

\bibitem[Lee and Cha, 2016]{LeeCha2016}
Lee, H. and Cha, J.~H. (2016).
\newblock New stochastic models for preventive maintenance and maintenance
  optimization.
\newblock {\em European Journal of Operational Research}, 255(1):80--90.

\bibitem[Liao et~al., 2006]{LiaElsCha2006}
Liao, H., Elsayed, E.~A., and Chan, L.-Y. (2006).
\newblock Maintenance of continuously monitored degrading systems.
\newblock {\em European Journal of Operational Research}, 175(2):821--835.

\bibitem[Liao et~al., 2010]{LiaPanXi2010}
Liao, W., Pan, E., and Xi, L. (2010).
\newblock Preventive maintenance scheduling for repairable system with
  deterioration.
\newblock {\em Journal of Intelligent Manufacturing}, 21(6):875--884.

\bibitem[Liu et~al., 2017]{liuWuXie2017}
Liu, B., Wu, S., Xie, M., and Kuo, W. (2017).
\newblock A condition-based maintenance policy for degrading systems with
  age-and state-dependent operating cost.
\newblock {\em European Journal of Operational Research}, 263(3):879--887.

\bibitem[Liu et~al., 2018]{liuChenJiang2018}
Liu, Y., Chen, Y., and Jiang, T. (2018).
\newblock On sequence planning for selective maintenance of multi-state systems
  under stochastic maintenance durations.
\newblock {\em European Journal of Operational Research}, 268(1):113--127.

\bibitem[Liu and Huang, 2010]{LiuHua2010b}
Liu, Y. and Huang, H.-Z. (2010).
\newblock Optimal selective maintenance strategy for multi-state systems under
  imperfect maintenance.
\newblock {\em IEEE Transactions on Reliability}, 59(2):356--367.

\bibitem[Meier-Hirmer et~al., 2009]{MeiRibSou2009}
Meier-Hirmer, C., Riboulet, G., Sourget, F., and Roussignol, M. (2009).
\newblock Maintenance optimization for a system with a gamma deterioration
  process and intervention delay: application to track maintenance.
\newblock {\em Proceedings of the Institution of Mechanical Engineers, Part O:
  Journal of Risk and Reliability}, 223(3):189--198.

\bibitem[Mobley et~al., 2008]{MobHigWik2008}
Mobley, R., Higgins, L., and Wikoff, D. (2008).
\newblock {\em Maintenance engineering handbook}.
\newblock McGraw-Hill,.

\bibitem[Moghaddam, 2013]{Mog2013}
Moghaddam, K.~S. (2013).
\newblock Multi-objective preventive maintenance and replacement scheduling in
  a manufacturing system using goal programming.
\newblock {\em International Journal of Production Economics}, 146(2):704--716.

\bibitem[Moghaddam and Usher, 2011]{MogUsh2011}
Moghaddam, K.~S. and Usher, J.~S. (2011).
\newblock Preventive maintenance and replacement scheduling for repairable and
  maintainable systems using dynamic programming.
\newblock {\em Computers \& Industrial Engineering}, 60(4):654--665.

\bibitem[Mor and Mosheiov, 2015]{MorMos2015}
Mor, B. and Mosheiov, G. (2015).
\newblock Scheduling a deteriorating maintenance activity and due-window
  assignment.
\newblock {\em Computers \& Operations Research}, 57:33--40.

\bibitem[Mosheiov and Sarig, 2009]{MosSar2009}
Mosheiov, G. and Sarig, A. (2009).
\newblock A note: Simple heuristics for scheduling a maintenance activity on
  unrelated machines.
\newblock {\em Computers \& Operations Research}, 36(10):2759--2762.

\bibitem[Nakagawa and Yasui, 1987]{NakYas1987}
Nakagawa, T. and Yasui, K. (1987).
\newblock Optimum policies for a system with imperfect maintenance.
\newblock {\em IEEE Transactions on Reliability}, 36(5):631--633.

\bibitem[Nash, 1973]{Nash}
Nash, P. (1973).
\newblock {\em Optimal Allocation of Resources Between Research Projects}.
\newblock PhD thesis, Cambridge University.

\bibitem[Nguyen et~al., 2017]{NguDijFou2017}
Nguyen, D., Dijoux, Y., and Fouladirad, M. (2017).
\newblock Analytical properties of an imperfect repair model and application in
  preventive maintenance scheduling.
\newblock {\em European Journal of Operational Research}, 256(2):439--453.

\bibitem[Oyarbide-Zubillaga et~al., 2008]{OyaGotSan2008}
Oyarbide-Zubillaga, A., Goti, A., and Sanchez, A. (2008).
\newblock Preventive maintenance optimisation of multi-equipment manufacturing
  systems by combining discrete event simulation and multi-objective
  evolutionary algorithms.
\newblock {\em Production Planning \& Control}, 19(4):342--355.

\bibitem[Pandelis and Teneketzis, 1999]{PanTen}
Pandelis, D.~G. and Teneketzis, D. (1999).
\newblock On the optimality of the gittins index rule for multi-armed bandits
  with multiple plays.
\newblock {\em Mathematical Methods of Operations Research}, 50:449--461.

\bibitem[Papadimitriou and Tsitsiklis, 1999]{PapaTsi}
Papadimitriou, C. and Tsitsiklis, J. (1999).
\newblock The complexity of optimal queueing network control.
\newblock {\em Math. Oper. Res.}, 24:293--305.

\bibitem[Percy, 2008]{Percy2008}
Percy, D.~F. (2008).
\newblock Preventive maintenance models for complex systems.
\newblock In Kobbacy, K. and Murthy, P., editors, {\em Complex System
  Maintenance Handbook}, pages 179--207. Springer.

\bibitem[Pham and Wang, 1996]{PhaWan1996}
Pham, H. and Wang, H. (1996).
\newblock Imperfect maintenance.
\newblock {\em European journal of operational research}, 94(3):425--438.

\bibitem[Poppe et~al., 2018]{PopBouLam2018}
Poppe, J., Boute, R.~N., and Lambrecht, M.~R. (2018).
\newblock A hybrid condition-based maintenance policy for continuously
  monitored components with two degradation thresholds.
\newblock {\em European Journal of Operational Research}, 268(2):515--532.

\bibitem[Puterman, 1994]{Puterman}
Puterman, M.~L. (1994).
\newblock {\em Markov Decision Processes: Discrete and Stochastic Dynamic
  Programming}.
\newblock Wiley, New York.

\bibitem[Ross, 1983]{Ross}
Ross, S. (1983).
\newblock {\em Introduction to Stochastic Dynamic Programming}.
\newblock Academic Press, Florida, USA.

\bibitem[Ruiz et~al., 2007]{RuiGarMar2007}
Ruiz, R., Garc{\'\i}a-D{\'\i}az, J.~C., and Maroto, C. (2007).
\newblock Considering scheduling and preventive maintenance in the flowshop
  sequencing problem.
\newblock {\em Computers \& Operations Research}, 34(11):3314--3330.

\bibitem[Ruiz-Hern{\'a}ndez, 2007]{Ruiz}
Ruiz-Hern{\'a}ndez, D. (2007).
\newblock {\em Indexable Restless Bandits. Index Policies for Some Stochastic
  Scheduling and Dynamic Allocation Problems.}
\newblock Verlag Dr. M{\"u}ller, Saarbr{\"u}cken, Germany.

\bibitem[Seif and Andrew, 2018]{SeiYu2018}
Seif, J. and Andrew, J.~Y. (2018).
\newblock An extensive operations and maintenance planning problem with an
  efficient solution method.
\newblock {\em Computers \& Operations Research}, 95:151--162.

\bibitem[Tao et~al., 2014]{TaoXiaXi2014}
Tao, X., Xia, T., and Xi, L. (2014).
\newblock Opportunistic preventive maintenance scheduling based on theory of
  constraints.
\newblock In {\em IIE Annual Conference. Proceedings}, page 230. Institute of
  Industrial Engineers-Publisher.

\bibitem[Valdez-Flores and Feldman, 1989]{ValFel1989}
Valdez-Flores, C. and Feldman, R.~M. (1989).
\newblock A survey of preventive maintenance models for stochastically
  deteriorating single-unit systems.
\newblock {\em Naval Research Logistics (NRL)}, 36(4):419--446.

\bibitem[Wang and Sheu, 2003]{WanShe2003}
Wang, C.-H. and Sheu, S.-H. (2003).
\newblock Determining the optimal production--maintenance policy with
  inspection errors: using a markov chain.
\newblock {\em Computers \& Operations Research}, 30(1):1--17.

\bibitem[Wang, 2002]{Wan2002}
Wang, H. (2002).
\newblock A survey of maintenance policies of deteriorating systems.
\newblock {\em European journal of operational research}, 139(3):469--489.

\bibitem[Wang and Pham, 2006]{WanPha2006b}
Wang, H. and Pham, H. (2006).
\newblock {\em Reliability and optimal maintenance}.
\newblock Springer Science \& Business Media.

\bibitem[Wang and Wei, 2011]{WanWei2011}
Wang, J.-B. and Wei, C.-M. (2011).
\newblock Parallel machine scheduling with a deteriorating maintenance activity
  and total absolute differences penalties.
\newblock {\em Applied Mathematics and Computation}, 217(20):8093--8099.

\bibitem[Weber and Weiss, 1990]{WebWei}
Weber, R. and Weiss, G. (1990).
\newblock On an index policy for restless bandits.
\newblock {\em J. Appl. Prob.}, 27:637--648.

\bibitem[Whittle, 1980]{Whit1}
Whittle, P. (1980).
\newblock Multi-armed bandits and the {Gittins} index.
\newblock {\em Journal of the Royal Statistical Society, Series B},
  42:143--149.

\bibitem[Whittle, 1981]{Whit2}
Whittle, P. (1981).
\newblock Arm-acquiring bandits.
\newblock {\em Annals of Probability}, 9:284--292.

\bibitem[Whittle, 1988]{Whit3}
Whittle, P. (1988).
\newblock Restless bandits: Activity allocation in a changing world.
\newblock {\em Journal of Applied Probability}, 25A:287--298.

\bibitem[Whittle, 1996]{WhitOC}
Whittle, P. (1996).
\newblock {\em Optimal Control, Basics and Beyond}.
\newblock John Wiley, Chichester.

\bibitem[Zhu et~al., 2015]{ZhuPenVan2015}
Zhu, Q., Peng, H., and van Houtum, G.-J. (2015).
\newblock A condition-based maintenance policy for multi-component systems with
  a high maintenance setup cost.
\newblock {\em Or Spectrum}, 37(4):1007--1035.

\end{thebibliography}

\end{document}